\documentclass[12pt,a4paper]{article}
\usepackage{epsfig}
\title{On the strong coupling region \\
in quantum matrix string theory}
 
\author{{\sc Shozo
Uehara}\footnote{e-mail: uehara@eken.phys.nagoya-u.ac.jp}{} ~and
{\sc Satoshi Yamada}\footnote{e-mail:
yamada@eken.phys.nagoya-u.ac.jp}\vspace{4mm}\\
{\sl Department of Physics, Nagoya University} \\
{\sl Chikusa-ku, Nagoya 464-8602, Japan}}
\date{}

\makeatletter

\@addtoreset{equation}{section}
\renewcommand{\thefigure}{\@arabic\c@figure}
\makeatother

\newcommand{\nn}{\nonumber\\}
\newcommand{\calL}{\mathcal{L}}
\newcommand{\calD}{\mathcal{D}}
\newcommand{\diag}{\rm diag}
\newcommand{\odiag}{\mbox{\scriptsize off-diag}}
\newcommand{\ptau}{\partial_\tau}
\newcommand{\pth}{\partial_\theta}
\newcommand{\ptaup}{\partial_{\tau'}}
\newcommand{\pthp}{\partial_{\theta'}}
\newcommand{\vev}[1]{\langle #1 \rangle}

\begin{document}
\maketitle
\vspace{-80mm}
\begin{flushright}
DPNU-02-24\\
hep-th/0207209\\
July 2002
\end{flushright}
\vspace{50mm}
\begin{abstract}
We study the behavior of matrix string theory in the strong coupling
region, where matrix strings reduce to discrete light-cone type IIA
superstrings except at the usual string-interaction points.
In the large $N$ limit, this reduction corresponds to the
double-dimensional reduction from wrapped supermembranes on $R^{10}
\times S^1$ to type IIA superstrings on $R^{10}$ in the light-cone
gauge. Such reductions were shown classically, while they are not
obvious quantum mechanically.
Recently, Sekino and Yoneya analyzed the double-dimensional reduction
of the wrapped supermembrane quantum mechanically 
to one-loop order in the strong
coupling expansion.
We analyze the problem in matrix string theory by using the same
expansion. At the one-loop level, the quantum corrections cancel out
as was presented by them.
However, at the two-loop level we find that the quantum corrections
cancel out only for the leading terms in the large $N$.
\end{abstract}

\section{Introduction}
It is believed that the supermembrane in eleven dimensions
\cite{HLP,BST} plays an important role to understand the fundamental
degrees of freedom in M-theory which is a unified description of the
various superstring theories.
Actually, it was shown that the supermembrane in eleven dimensions
is related to type IIA superstring in ten dimensions by means
of the classical double-dimensional reduction \cite{DHIS}.
The procedure is the following: (i) Consider the target space of
$R^{10}\times S^1$. (ii) Set the compactified coordinate (with radius
$L$) proportional to one of the spatial coordinates of the world volume,
which we call $\rho$ coordinate. (iii) Simply ignore the infinite
tower of the Kaluza-Klein (non-zero) modes.
However, it is not obvious whether such a reduction is justified
also in quantum theory.
Actually, it was pointed out that the other set of zero-mode states
which are independent of the other spatial coordinate of the world
volume, which we call $\sigma$ coordinate, do not decouple even in the
zero-radius limit ($L\to 0$) \cite{Rus}.
Hence, whether the Kaluza-Klein modes along the compactified $\rho$
direction are suppressed quantum mechanically seems to be a subtle
question.

Sekino and Yoneya analyzed the double-dimensional reduction quantum
mechanically with the light-cone supermembrane action in the appendix
of their paper \cite{SY}.
Contrary to the classical treatments, they kept the Kaluza-Klein modes
associated with the $\rho$ coordinate in the wrapped supermembrane
theory on the target space $R^{10}\times S^1$ and they integrated them
out by using the perturbative expansion with respect to the radius
$L$.
Since the gauge coupling satisfies $g\sim 1/L$ in the wrapped
supermembrane theory, the expansion can be regarded as the strong
coupling expansion.
They calculated the effective action for the zero modes along the
$\rho$ direction to the one-loop order of $O(L^2)$ by integrating out
the Kaluza-Klein modes. They found that the quantum
corrections cancel out and the effective action agrees with the
classical (free) action of type IIA superstring except at the points
where the usual string interactions could occur.
However, as is emphasized in their paper \cite{SY}, the strong
coupling expansion does not give a rigorous proof of the quantum
double-dimensional reduction.
The free parts of the Kaluza-Klein modes in the strong coupling
expansion have no derivatives and it leads to the propagators which
are proportional to the two-dimensional $\delta$-function,
$\delta^{(2)}(\xi) \equiv \delta(\tau)\delta(\sigma)$.
Thus, the loop diagrams suffer from the ultraviolet divergences of
$\delta^{(2)}(0)$ type, and we need a regularization for a rigorous
treatment.
However it is very difficult to find a suitable regularization which
respects symmetries (e.g., supersymmetry and gauge symmetry), and
hence the strong coupling expansion is not defined rigorously.
In this sense, they gave a formal argument for the vanishing of the
one-loop corrections of $O(L^2)$ by demonstrating that the
coefficients of $\delta^{(2)}(0)$ coming from both bosonic and
fermionic degrees of freedom cancel out.

The purpose of this paper is essentially to extend their (formal)
calculations to the two-loop order of $O(L^2)$.
However, the naive extension is not straightforward because at the
two-loop level, even the coefficients of the $\delta^{(2)}(0)$ diverge
due to the contribution of the infinite Kaluza-Klein towers.
Thus, we need another regularization for the summation over the
infinite tower of the Kaluza-Klein modes at the two-loop level.
Contrary to the case of the divergence of $\delta^{(2)}(0)$ itself,
it is relatively easy to find a regularization (which respects
symmetries) for the divergence of the coefficients due to the
infinite Kaluza-Klein tower along the compactified $\rho$ direction.
In fact, we know the matrix regularization of the supermembrane on
$R^{11}$ in the light-cone gauge \cite{dHN} and also that of the
wrapped supermembrane on $R^{10} \times S^1$ in the light-cone gauge
\cite{SY}\footnote{In Ref.\cite{SY}, the quantum mechanical study on
the double-dimensional reduction is discussed in appendix A, and in
the body of the paper, the correspondence of the degrees of freedom in
the wrapped supermembrane theory with those in matrix string theory is
discussed in detail.}.
The former is called Matrix theory \cite{BFSS} which was proposed to
be a non-perturbative formulation of light-cone quantized M-theory in
the large $N$ limit and the latter is called matrix string theory
\cite{Mot,DVV} which will be a non-perturbative formulation of
light-cone quantized type IIA superstring theory in the large $N$
limit.
Furthermore, even at finite $N$, Matrix and matrix string theories
are conjectured to be non-perturbative formulations of discrete
light-cone quantized (DLCQ)\footnote{In this paper we use a
convention of the light-cone coordinates such that $x^{\pm}=(x^{0}\pm
x^{10})/\sqrt 2$. Furthermore, $x^-$ is compactified on $S^1$ with
radius $R$ in DLCQ.}
M-theory and type IIA superstring theory, respectively
\cite{Sus,Sei,Sen}.
Thus, in this paper
we consider matrix string theory and study whether the reduction from
matrix strings to discrete light-cone type IIA superstrings is
justified quantum mechanically.
According to the correspondence of the wrapped supermembrane with
matrix string \cite{SY}, the zero modes along the $\rho$ direction,
i.e., type IIA superstring degrees of freedom in the wrapped
supermembrane theory, are mapped to the diagonal elements in matrix
string theory.
Hence, in this paper we study whether the reduction from matrix
strings to the diagonal elements of strings is justified quantum
mechanically to the two-loop order of $O(L^2)$, except at the points
where the strings could interact usually.

The plan of this paper is as follows.
In the next section, we review the correspondence
of the wrapped supermembranes on $R^{10} \times S^1$ with matrix
strings.
In section \ref{SCEinMST} we discuss the strong coupling expansion in
matrix string theory. By using the expansion in the path-integral
formula, we integrate out the off-diagonal matrix elements to the
two-loop order of $O(L^2)$.  We obtain the effective action for the
diagonal matrix elements and study whether the reduction from matrix
strings to the diagonal elements is quantum mechanically justified or
not. Section \ref{sec:CandD} is devoted to our conclusion and
discussion. In appendix A, we put the explicit expressions of the
interaction parts of the action.

\section{From wrapped supermembrane to matrix string\label{sec:MtoS}}
Our starting point is the following light-cone gauge fixed
supermembrane action on the target space $R^{11}$ \cite{dHN},
\begin{eqnarray}
  S\!&=\!&LT\int d\tau \!\int_0^{2\pi}\!d\sigma d\rho \left[
    \frac{1}{2}(D_{\tau}X^i)^2 -\frac{1}{4L^2}\{ X^i,X^j\}^2
    +i\psi^T D_{\tau}\psi +\frac{i}{L}\psi^T\gamma^i\{ X^i,\psi\}
	\right]\label{action1}\!,\\
  &&D_{\tau} X^i =\ptau X^i -\frac{1}{L}\{A,X^i\},\\
  &&D_{\tau} \psi =\ptau \psi -\frac{1}{L}\{A,\psi\},\\
  &&\{A,B\} = \partial_{\sigma}A\,\partial_{\rho}B -
	\partial_{\rho}A\, \partial_{\sigma}B ,
\end{eqnarray}
where the indices $i,j$ run through $1,2,\cdots,9$,
the spinor $\psi$ has sixteen real components\footnote{We use the real
and symmetric representation for the gamma matrices $\gamma^i$, which
satisfy $\{\gamma^i,\gamma^j\}=2\delta^{ij}$.}
and $T$ is the membrane tension.
At this stage, $L$ is an arbitrary length parameter of no physical
meaning. This action is invariant under the gauge transformation,
\begin{eqnarray}
  \delta A&=&\ptau \Lambda + \frac{1}{L}\{\Lambda,A\},\\
  \delta X^i&=& \frac{1}{L}\{\Lambda,X^i\},\\
  \delta \psi&=& \frac{1}{L}\{\Lambda,\psi\}.
\end{eqnarray}
This gauge transformation generates the area-preserving diffeomorphism
on the world volume.
In the $A=0$ gauge, the Gauss law constraint derived from the action
(\ref{action1}) is given by
\begin{equation}
    \{\ptau X^i, X^i\} +i\{\psi^T, \psi \}=0.\label{Gauss}
\end{equation}
This constraint is originally the integrability condition for the
equations determining the light-cone coordinate $X^-$,
\begin{equation}
  \frac{1}{LT}P^+\partial_{\hat{\sigma}}X^-
	=\partial_{\hat{\sigma}}X^i\ptau X^i
	+i\psi^T\partial_{\hat{\sigma}} \psi\,,
\end{equation}
where $\hat{\sigma}=(\sigma,\rho)$ and this equation is locally
equivalent to eq.(\ref{Gauss}). Note that the light-cone
momentum (density), $P^+=LT \ptau X^+$, is constant on the
world volume.
When the spatial surface of the supermembrane has a non-trivial
topology, we have to impose further the global constraints.
Actually, in the case of the toroidal supermembrane\footnote{In this
paper we consider this case only.}, the global constraints are given
by
\begin{equation}
  \int_0^{2\pi}d \sigma (\partial_{\sigma}X^i\ptau X^i
	+i\psi^T\partial_{\sigma} \psi)
	=\int_0^{2\pi}d \rho (\partial_{\rho}X^i\ptau X^i
	+i\psi^T\partial_{\rho} \psi)=0.\label{global}
\end{equation}

Now, we consider the wrapped supermembrane theory on the target space
$R^{10}\times S^1$ and discuss the correspondence of the wrapped
supermembrane with matrix string \cite{SY}.
We take the $X^9$ direction as the $S^1$ and identify the radius with
the above parameter $L$,
\begin{eqnarray}
	X^9=L\rho +Y.\label{L}
\end{eqnarray}
Thus $L$ has the physical meaning of a radius of the $X^9$ direction
which is regarded as the ``eleventh'' direction in M-theory.
Substituting eq.(\ref{L}) into eq.(\ref{action1}), we obtain the
following light-cone gauge fixed supermembrane action on
$R^{10}\times S^1$,
\begin{eqnarray}
  S&=&LT\int d\tau \int_0^{2\pi}d\sigma d\rho \left[
	\frac{1}{2}F_{\tau\sigma}^2 +\frac{1}{2}(D_{\tau}X^k)^2
	-\frac{1}{2}(D_{\sigma}X^k)^2 -\frac{1}{4L^2}\{ X^k,X^l\}^2
	\right.\nn
  &&\hspace{26ex} \left.+i\psi^T D_{\tau}\psi -i\psi^T \gamma^9
	D_{\sigma}\psi +\frac{i}{L}\psi^T\gamma^k\{ X^k,\psi\}
	\right], \label{action}\\
  &&F_{\tau\sigma}=\ptau  Y - \partial_{\sigma}A -
	\frac{1}{L}\{A,Y\},\\
  &&D_{\sigma}X^k=\partial_{\sigma}X^k-\frac{1}{L}\{Y,X^k\},\\
  &&D_{\sigma}\psi=\partial_{\sigma}\psi-\frac{1}{L}\{Y,\psi\},
\end{eqnarray}
where the indices $k,l$ run through $1,2,\cdots,8$.
This is also an action of the gauge theory of the area-preserving
diffeomorphism, where the gauge coupling $g\sim 1/L$.
The gauge transformations are as follows,
\begin{eqnarray}
  \delta A&=&\ptau  \Lambda + \frac{1}{L}\{\Lambda,A\},\\
  \delta Y&=&\partial_{\sigma} \Lambda + \frac{1}{L}\{\Lambda,Y\},\\
  \delta X^k&=& \frac{1}{L}\{\Lambda,X^k\},\\
  \delta \psi&=& \frac{1}{L}\{\Lambda,\psi\}.
\end{eqnarray}
Furthermore, substituting eq.(\ref{L}) into eqs.(\ref{global}),
we have the global constraints
\begin{eqnarray}
 && \int_0^{2\pi}d \sigma (\partial_{\sigma}Y\ptau Y
	+\partial_{\sigma}X^k\ptau X^k
	+i\psi^T\partial_{\sigma} \psi)=0,\label{global1}\\
 && \int_0^{2\pi}d \rho (L\ptau Y
	+\partial_{\rho}Y\ptau Y
	+\partial_{\rho}X^k\ptau X^k
	+i\psi^T\partial_{\rho} \psi)=0.\label{global2}
\end{eqnarray}
In Ref.\cite{SY}, the infinite dimensional gauge group of the
area-preserving diffeomorphism in eq.(\ref{action}) was regularized by
the finite dimensional group $U(N)$ and it was shown that the 
matrix-regularized form of the action (\ref{action}) agrees with that of
matrix string theory,
\begin{eqnarray}
  S&=&LT\int d\tau \int_0^{2\pi}d\theta \,tr\left[
	\frac{1}{2}F_{\tau\theta}^2 +\frac{1}{2}(D_{\tau}X^k)^2
	-\frac{1}{2}(D_{\theta}X^k)^2 +\frac{1}{4L^2}[ X^k,X^l]^2
	\right.\label{MSTaction}\nn
  &&\hspace{26ex} \left.+i\psi^T D_{\tau}\psi -i\psi^T \gamma^9
	D_{\theta}\psi -\frac{1}{L}\psi^T\gamma^k[X^k,\psi]
	\right],\label{action2}\\
  &&F_{\tau\theta}=\ptau  Y - \pth A - \frac{i}{L}[A,Y],\\
  &&D_{\tau}X^k=\ptau X^k-\frac{i}{L}[A,X^k],\\
  &&D_{\theta}X^k=\pth X^k-\frac{i}{L}[Y,X^k]\,,
\end{eqnarray}
where each element of the matrices is a function of $\tau$ and
$\theta$.
Note that the action (\ref{MSTaction}) can be derived from Matrix
theory action by combining T- and S-dualities with the flipping of the
compactified direction from eleventh to ninth \cite{Mot,DVV}.
The $U(N)$ gauge transformations of the action (\ref{action2}) are as
follows,
\begin{eqnarray}
  \delta A&=&\ptau  \Lambda + \frac{i}{L}[\Lambda,A],\\
  \delta Y&=&\pth  \Lambda + \frac{i}{L}[\Lambda,Y],\\
  \delta X^k&=& \frac{i}{L}[\Lambda,X^k],\\
  \delta \psi&=& \frac{i}{L}[\Lambda,\psi].
\end{eqnarray}
In the correspondence between the actions (\ref{action}) and
(\ref{action2}), the zero-modes along the $\rho$ direction in the
wrapped supermembrane are mapped to the diagonal elements of matrix
string and the Kaluza-Klein modes are mapped to the
off-diagonal elements \cite{SY}.
Here, we should notice that in the matrix regularization
of the wrapped supermembrane on $R^{10} \times S^1$,
we have no obvious counterparts of the global constraints
(\ref{global1}) and (\ref{global2}), because the (matrix-regularized)
Gauss law constraint, which is derived from eq.(\ref{action2}),
cannot be manifestly interpreted as the integrability
condition\footnote{Of course, also in the matrix regularization
of the supermembrane on $R^{11}$ \cite{dHN}, i.e., Matrix theory,
we have no obvious counterparts of the global constraints
(\ref{global}) due to the same reason.}.
Furthermore, in the standard derivation \cite{Mot,DVV} of matrix
string theory based on Seiberg and Sen's arguments \cite{Sei,Sen}
and the compactification prescription of Taylor \cite{Tay},
such global constraints do not appear naturally\footnote{In
Ref.\cite{Mot}, the origin of the level-matching condition in matrix
string theory was discussed.}.

The classical double-dimensional reduction is to assume that the
Kaluza-Klein modes along the $\rho$ direction of every field are
zero. Then the action (\ref{action}) reduces to
\begin{equation}
  S=2\pi LT\int d\tau \int_0^{2\pi}d\sigma \left[
	\frac{1}{2}(\ptau X^k)^2
	-\frac{1}{2}(\partial_{\sigma}X^k)^2 +i\psi^T
	\ptau \psi -i\psi^T \gamma^9
	\partial_{\sigma}\psi \right],
\end{equation}
where, for simplicity, we also set the zero modes of $A$ and $Y$
fields to zero.
With the identification $2\pi LT =1/2\pi\alpha'$, which is kept finite
in the $L\to 0$ limit, this action agrees with the light-cone type IIA
superstring action in the Green-Schwarz formalism.
In the matrix-regularized action (\ref{action2}),
such a classical double-dimensional reduction corresponds to the
assumption that the off-diagonal elements of every matrix are
zero,
\begin{equation}
  X^k=\left(\begin{array}{cccc}
	x^k_1&&&\lower1ex\hbox{\Large0}\\
	&x^k_2&&\\
	&&\ddots&\\
	\hbox{\Large0}&&&x^k_N
	\end{array}\right),
  \quad\psi=\left(\begin{array}{cccc}
	\psi_1&&&\lower1ex\hbox{\Large0}\\
	&\psi_2&&\\
	&&\ddots&\\
	\hbox{\Large0}&&&\psi_N
	\end{array}\right)\,.\label{diag}
\end{equation}
Then the action reduces to the DLCQ type IIA superstring action
in the light-cone momentum $p^+=N/R$ sector.
Depending on the boundary conditions with respect to  $\theta$,
the diagonal elements  $x^k_a(\theta)\,\,(a=1,\cdots,N)$ in the matrix
(\ref{diag}) describe one or more separate strings. For example, the
boundary conditions $x^k_a(\theta +2\pi)=x^k_a(\theta)$ correspond to
$N$ string bits having $p^+=1/R$, which are regarded as the minimal
length strings in DLCQ.
On the other hand, a string of maximal length having  $p^+=N/R$
is described by the boundary condition $x^k_a(\theta
+2\pi)=x^k_{a+1}(\theta), \,x^k_{N+1}(\theta)=x^k_{1}(\theta)$.

It is expected that the above reductions are justified also in quantum
theory\footnote{To be precise, since it is not obvious whether the
quantum supermembranes can be the degrees of freedom in M-theory,
there may be no logical reasons for the expectation that the reduction
from the wrapped supermembranes to type IIA superstrings is justified
also in quantum theory. However, in the context of Matrix and matrix
string theories, the quantum matrix-regularized supermembranes are the
degrees of freedom in DLCQ M-theory.
Hence, it is expected that the reduction from matrix strings to DLCQ
type IIA strings is justified also in quantum theory.}.
However, as was discussed in Refs.\cite{Rus} and
\cite{SY}\footnote{See also Ref.\cite{HMS} for related discussions.},
the justification is not so simple. In particular, in the appendix of
Ref.\cite{SY}, the quantum double-dimensional reduction of the wrapped
supermembrane (\ref{action}) was analyzed for the small radius $L$,
which corresponds to the strong gauge coupling $g\sim 1/L$ in the
wrapped supermembrane theory and also to the weak string coupling
$g_s\sim L/\sqrt{\alpha'}$ in type IIA superstring theory.
Concretely, by using the perturbative expansion with respect to $L$ in
the path-integral formula, the Kaluza-Klein modes along the $\rho$
direction were integrated out to the one-loop order of $O(L^2)$,
and it was found that the effective action for the zero modes
agrees with the classical (free) action of the type IIA superstring
except at the points where perturbative interactions would occur by
joining or splitting of strings.
That result is consistent with the expectation that the wrapped
supermembrane theory in the region of small radius $L$ agrees with
the perturbative type IIA superstring theory.
In the next section, we analyze the quantum reduction of matrix string
(\ref{action2}) to the diagonal elements for small radius $L$.
That is, by using the same perturbative expansion in the
path-integral formula, we integrate out the off-diagonal matrix
elements to the two-loop order of $O(L^2)$ and study whether the
effective action for the diagonal matrix elements agrees with the
classical (free) action of the DLCQ type IIA superstring except at the
points where perturbative interactions would occur by joining or
splitting of DLCQ strings.

\section{Strong coupling expansion in matrix string
theory\label{SCEinMST}}
\subsection{Path-integral formula}
To begin with, we decompose every $N\times N$ hermite matrix in
eq.(\ref{action2}) into the diagonal and off-diagonal parts as
follows,
\begin{eqnarray}
	A&\to& a+A,\label{a1}\\
	Y&\to& y+Y,\\
	X^k&\to& x^k+X^k,\\
	\psi&\to& \psi+\Psi,\label{psi1}
\end{eqnarray}
where $a,y,x^k$ and $\psi$ are the diagonal parts;
\begin{eqnarray}
  a=(a_a)=\left(\begin{array}{cccc}
	a_1&&&\lower1ex\hbox{\Large 0}\\
	&a_2&&\\
	&&\ddots&\\
	\hbox{\Large 0}&&&a_N
	\end{array}
	\right),\,\,
  y=(y_a)=\left(\begin{array}{cccc}
	y_1&&&\lower1ex\hbox{\Large 0}\\
	&y_2&&\\
	&&\ddots&\\
	\hbox{\Large 0}&&&y_N
	\end{array}
	\right),\nn
  x^k=(x^k_a)=\left(\begin{array}{cccc}
	x^k_1&&&\lower1ex\hbox{\Large 0}\\
	&x^k_2&&\\
	&&\ddots&\\
	\hbox{\Large 0}&&&x^k_N
	\end{array}
	\right),\,\,
  \psi=(\psi_a)=\left(\begin{array}{cccc}
	\psi_1&&&\lower1ex\hbox{\Large 0}\\
	&\psi_2&&\\
	&&\ddots&\\
	\hbox{\Large 0}&&&\psi_N
	\end{array}
	\right),\label{diag2}
\end{eqnarray}
and $A,Y,X^k$ and $\Psi$ are the off-diagonal parts,
\begin{eqnarray}
  A&=&\Big(A_{ab}\Big)=\left(\begin{array}{cccc}
	0&A_{12}&\cdots&A_{1N}\\
	A_{21}&0&\cdots&A_{2N}\\
	\vdots&&\ddots&\\
	A_{N1}&A_{N2}&\cdots&0
	\end{array}\right),\quad
  Y=\Big(Y_{ab}\Big)=\left(\begin{array}{cccc}
	0&Y_{12}&\cdots&Y_{1N}\\
	Y_{21}&0&\cdots&Y_{2N}\\
	\vdots&&\ddots&\\
	Y_{N1}&Y_{N2}&\cdots&0
	\end{array}\right),\nn
  X^k&=&\Big(X^k_{ab}\Big)=\left(\begin{array}{cccc}
	0&X^k_{12}&\cdots&X^k_{1N}\\
	X^k_{21}&0&\cdots&X^k_{2N}\\
	\vdots&&\ddots&\\
	X^k_{N1}&X^k_{N2}&\cdots&0
	\end{array}\right),\nn
  \Psi&=&\Big(\Psi_{ab}\Big)=\left(\begin{array}{cccc}
	0&\Psi_{12}&\cdots&\Psi_{1N}\\
	\Psi_{21}&0&\cdots&\Psi_{2N}\\
	\vdots&&\ddots&\\
	\Psi_{N1}&\Psi_{N2}&\cdots&0
\end{array}
\right).\label{off-diag}
\end{eqnarray}
After the replacement of eqs.(\ref{a1})-(\ref{psi1}),
the action (\ref{action2}) becomes
\begin{eqnarray}
  S&=&LT\int d\tau \int_0^{2\pi}d\theta(\calL_B+\calL_F),\label{S}\\
  \calL_B &=&tr\left[\,\frac{1}{2}(\partial_\theta a)^2
	-\partial_\theta a \ptau  y
	+ \frac{1}{2}(\ptau  y)^2
	+ \frac{1}{2}(\ptau  x^k)^2
	- \frac{1}{2}(\partial_\theta x^k)^2\right.\\
  &&\hspace{1cm}-\frac{1}{2L^2}([a,Y]+[A,y])^2
	-\frac{1}{2L^2}([a,X^k]+[A,x^k])^2 \nn
  &&\hspace{1cm}+\frac{1}{2L^2}([y,X^k]+[Y,x^k])^2
      +\frac{1}{4L^2}([x^k,X^l]+[X^k,x^l])^2\nn
  &&\hspace{1cm}-\frac{i}{L}(\ptau  Y-\partial_\theta A)
	([a,Y]+[A,y])-\frac{i}{L}
	(\ptau  y-\partial_\theta a)[A,Y]\nn
  &&\hspace{1cm}-\frac{i}{L}\ptau  X^k
	([a,X^k]+[A,x^k])-\frac{i}{L}\ptau  x^k[A,X^k]\nn
  &&\hspace{1cm}+\frac{i}{L}\partial_\theta X^k
	([y,X^k]+[Y,x^k])+\frac{i}{L}\partial_\theta x^k[Y,X^k]\nn
  &&\hspace{1cm}-\frac{1}{L^2}([a,Y]+[A,y])[A,Y]
	 -\frac{1}{L^2}([a,X^k]+[A,x^k])[A,X^k]\nn
  &&\hspace{1cm} +\frac{1}{L^2}([y,X^k]+[Y,x^k])[Y,X^k]
	 +\frac{1}{2L^2}([x^k,X^l]+[X^k,x^l])[X^k,X^l]\nn
  &&\hspace{1cm}+\frac{1}{2}(\ptau Y-\pth  A)^2
	+\frac{1}{2}(\ptau  X^k)^2
	-\frac{1}{2}(\pth X^k)^2\nn
  &&\hspace{1cm} -\frac{i}{L}\ptau  Y[A,Y]+
	\frac{i}{L}\pth  A[A,Y]
	-\frac{i}{L}\ptau  X^k[A,X^k]
	+\frac{i}{L}\pth  X^k[Y,X^k]\nn
  &&\hspace{1cm}\left.-\frac{1}{2L^2}[A,Y]^2 -\frac{1}{2L^2}[A,X^k]^2
	+\frac{1}{2L^2}[Y,X^k]^2
	+\frac{1}{4L^2}[X^k,X^l]^2\,\right],\nn
  \calL_F&=&tr\left[\,i\psi^T\ptau  \psi-i
	\psi^T\gamma^9\pth  \psi
	+\frac{1}{L} \Psi^T[a,\Psi]
	-\frac{1}{L}\Psi^T\gamma^9[y,\Psi]-
	\frac{1}{L}\Psi^T\gamma^k[x^k,\Psi]\right.\\
  &&\hspace{1cm} +\frac{2}{L}\psi^T[A,\Psi]-\frac{2}{L}\psi^T\gamma^9
	[Y,\Psi] -\frac{2}{L}\psi^T\gamma^k[X^k,\Psi]\nn
  &&\hspace{1cm} \left.+i\Psi^T\ptau \Psi
	 -i\Psi^T\gamma^9\pth \Psi
	+\frac{1}{L}\Psi^T[A,\Psi]- \frac{1}{L}\Psi^T\gamma^9[Y,\Psi]
	-\frac{1}{L}\Psi^T\gamma^k[X^k,\Psi]\right]\nonumber.
\end{eqnarray}
Furthermore, the gauge transformations are decomposed as
\begin{eqnarray}
  \delta a&=&\ptau \lambda
	+\frac{i}{L}[\Lambda,A]_{\diag}\,,\\
  \delta A&=&\ptau  \Lambda
	+ \frac{i}{L}([\lambda,A]+[\Lambda,a]
	+[\Lambda,A]_{\odiag})\,,\\
  \delta y&=&\pth  \lambda
	+ \frac{i}{L}[\Lambda,Y]_{\diag}\,,\\
  \delta Y&=&\pth  \Lambda
	+ \frac{i}{L}([\lambda,Y]+[\Lambda,y]
	+[\Lambda,Y]_{\odiag})\,,\\
  \delta x^k&=& \frac{i}{L}[\Lambda,X^k]_{\diag}\,,\\
  \delta X^k&=& \frac{i}{L}([\lambda,X^k]+[\Lambda,x^k]
	+[\Lambda,X^k]_{\odiag})\,,\\
  \delta \psi&=& \frac{i}{L}[\Lambda,\Psi]_{\diag}\,,\\
	\delta \Psi&=& \frac{i}{L}([\lambda,\Psi]+[\Lambda,\psi]
	+[\Lambda,\Psi]_{\odiag})\,,
\end{eqnarray}
where $\lambda$ and $\Lambda$ are the diagonal and off-diagonal
parts of the gauge function, respectively.
At this stage, we impose boundary conditions with respect to
$\theta$ on the diagonal matrix elements in eq.(\ref{diag2}).
Actually we choose such boundary conditions for the $N$ string
bits\footnote{In the large $N$ limit, $N$ string bits having $p^+=1/R$
do not correspond to the wrapped supermembrane directly. In fact, in
Ref.\cite{SY}, the correspondence of a long string having $p^+=N/R$
with the wrapped supermembrane was discussed. In this paper, however,
we study only the reduction from matrix strings to the $N$ string
bits for simplicity.} having $p^+=1/R$ as
\[
	a_a(\theta +2\pi)=a_a(\theta),\quad
	y_a(\theta +2\pi)=y_a(\theta),\quad
	x^k_a(\theta +2\pi)=x^k_a(\theta),\quad
	\psi_a(\theta +2\pi)=\psi_a(\theta).
\]
Then, it is natural that the off-diagonal matrix elements
in eq.(\ref{off-diag}) also satisfy the following boundary conditions,
\begin{eqnarray*}
 &&A_{ab}(\theta +2\pi)=A_{ab}(\theta),\quad
	Y_{ab}(\theta +2\pi)=Y_{ab}(\theta),\\
 &&X^k_{ab}(\theta +2\pi)=X^k_{ab}(\theta),\quad
	\Psi_{ab}(\theta +2\pi)=\Psi_{ab}(\theta).
\end{eqnarray*}

Our next task is to consider the path-integral formula of the action
(\ref{S}). The gauge conditions for the diagonal and off-diagonal
parts are chosen as follows,
\begin{eqnarray}
  &&a=y,\label{fixing1}\\
  &&\pth Y-\frac{i}{L}[y,Y]
	-\frac{i}{L}[x^k,X^k]+\frac{i}{L}[a,A]
	-\ptau A=0.\label{fixing2}
\end{eqnarray}
We proceed in the Landau gauge. Then, we obtain the path-integral
formula of the action (\ref{S}),
\begin{eqnarray}
 T&=&\int \calD y \calD x^k \calD\psi \calD c \calD\bar{c}
	\calD A \calD Y \calD X^k \calD \Psi \calD C \calD \bar{C}
	\calD B \,\exp\left[i(S+S_{gf}+S_{gh})\right],\label{PI}\\
  &&S+S_{gf}+S_{gh}=LT \int d\tau \int_0^{2\pi}
	d\theta (\calL_B +\calL_F +\calL_{gf}
	+\calL_{gh}),\label{Action}\\
  &&\calL_{gf}=tr\left[B\left(\pth Y-\frac{i}{L}[y,Y]
	-\frac{i}{L}[x^k,X^k]+\frac{i}{L}[y,A]
	-\ptau A\right)\right],\\
  &&\calL_{gh}=tr\left[i\bar{c}(\ptau -\pth )c
	-i\left\{\pth \bar{C}\left(\pth C
	-\frac{i}{L}[y,C]-\frac{i}{L}[Y,C]\right)\right.\right.\nn
  &&\hspace{2cm}-\frac{i}{L}[y,\bar{C}]\left(\pth C
	-\frac{i}{L}[y,C]-\frac{i}{L}[Y,C]\right)
	+\frac{1}{L^2}[\bar{C},Y]_{\diag}[C,Y]_{\diag}\nn
  &&\hspace{2cm}-\frac{1}{L^2}[x^k,\bar{C}]
	\left([x^k,C]+[X^k,C]\right)
	+\frac{1}{L^2}[\bar{C},X^k]_{\diag}[C,X^k]_{\diag}\nn
  &&\hspace{2cm}+\frac{i}{L}[y,\bar{C}]\left(\ptau C
	-\frac{i}{L}[y,C]-\frac{i}{L}[A,C]\right)
	-\frac{1}{L^2}[\bar{C},A]_{\diag}[C,A]_{\diag}\nn
  &&\hspace{2cm}\left.\left.-\ptau \bar{C}
	\left(\ptau C-\frac{i}{L}[y,C]
	-\frac{i}{L}[A,C]\right)\right\}\right],\label{ghost}
\end{eqnarray}
where the integration over $a$ is carried out by using the Landau
gauge condition for eq.(\ref{fixing1}).
Note that in eq.(\ref{ghost}), the coupling terms between the diagonal
part of the ghost (anti-ghost) and the off-diagonal part of the
anti-ghost (ghost), such as $\bar{C}c$ $(\bar{c}C)$, do not exist.
This is due to the Landau gauge for the gauge condition
eq.(\ref{fixing2}).
Now the off-diagonal parts are rescaled as \cite{SY}
\begin{equation}
  A\to LA, \, Y\to LY,\, X^k\to LX^k,\,
	 \Psi\to L^{1/2}\Psi,\, \bar{C}\to\bar{C},\,
	C\to L^2C,\, B\to B .
\end{equation}
Then the action (\ref{Action}) is given by
\begin{eqnarray}
  S+S_{gf}+S_{gh}&=&LT \int d\tau \int_0^{2\pi}
    d\theta\, (\calL^{string}+ \calL^B_0 +L\calL^B_1 + L^2\calL^B_2\nn
    &&\hspace{25ex}+\calL^F_0 +L^{1/2}\calL^F_{1/2} +L\calL^F_1),\nn
  \calL^{string}&=&tr\bigg[\frac{1}{2}(\ptau x^k)^2
	- \frac{1}{2}(\pth x^k)^2
	+\frac{1}{2}\left\{(\ptau -\pth)y\right\}^2\nn
  &&\hspace{2cm}+i\bar{c}(\ptau -\pth )c
	+ i\psi^T\ptau  \psi-i
	\psi^T\gamma^9\pth  \psi \bigg],\\
 \calL^B_0&=&tr\bigg[-\frac{1}{2}\,[x^k,A]^2+\frac{1}{2}\,[x^k,Y]^2
	+\frac{1}{2}\,[x^k,X^l]^2\nn
  &&\hspace{10ex}-\frac{1}{2}\left([y,Y]+[x^k,X^k]-[y,A]\right)^2\nn
  &&\hspace{10ex}-iB \left([y,Y]+[x^k,X^k]-[y,A]\right)
	+i[x^k,\bar{C}][x^k,C]\bigg],\label{freeb}\\
 \calL^B_1&=&tr\bigg[-i\ptau Y [y,Y] + 2i\ptau Y [y,A]-i\ptau A [y,Y]
	+ 2i\pth A [y,Y]\nn
  &&\hspace{1cm}- i\pth A[y,A]-i\pth Y[y,A]
	-i\ptau X^k[y,X^k]+2i\ptau X^k[x^k,A]\nn
  &&\hspace{1cm}-i\ptau A[x^k,X^k]+i\pth X^k[y,X^k]
	-2i\pth X^k[x^k,Y]
	+i\pth Y[x^k,X^k]\nn
  &&\hspace{1cm}-[y,Y][A,Y] +[y,A][A,Y]
	-[y,X^k][A,X^k]+[x^k,A][A,X^k]\nn
  &&\hspace{1cm}+[y,X^k][Y,X^k]-[x^k,Y][Y,X^k]+[x^k,X^l][X^k,X^l]\nn
  &&\hspace{1cm}+B\pth Y-B\ptau A\nn
  &&\hspace{1cm}-\pth \bar{C}[y,C]
	-[y,\bar{C}]\pth C+i[y,\bar{C}][Y,C]
	+i[x^k,\bar{C}][X^k,C]\nn
  &&\hspace{1cm}+ [y,\bar{C}]\,\ptau  C
	 -i[y,\bar{C}][A,C]+\ptau \bar{C}
	[y,C]\bigg],\label{1b}\\
  \calL^B_2&=&tr\bigg[\,\frac{1}{2}(\ptau  Y - \pth A)^2
	+\frac{1}{2}(\ptau  X^k)^2 -\frac{1}{2}(\pth  X^k)^2\nn
  &&\hspace{3ex}-i\ptau Y[A,Y] +i\pth A[A,Y]
	-i\ptau X^k[A,X^k] +i\pth X^k[Y,X^k]\nn
  &&\hspace{3ex}-\frac{1}{2}\,[A,Y]^2  -\frac{1}{2}\,[A,X^k]^2
	+ \frac{1}{2}\,[Y,X^k]^2+ \frac{1}{4}\,[X^k,X^l]^2\nn
  &&\hspace{3ex}-i\pth \bar{C}\pth C
	-\pth \bar{C}[Y,C] +i\ptau \bar{C}\ptau C
	+\ptau \bar{C}[A,C]\nn
  &&\hspace{3ex} +i[\bar{C},A]_{\diag}[C,A]_{\diag}
	-i[\bar{C},Y]_{\diag}[C,Y]_{\diag}\nn
  &&\hspace{30ex}-i[\bar{C},X^k]_{\diag}
	[C,X^k]_{\diag}\bigg],\label{2b}\\
 \calL^F_0&=&tr\left[\Psi^T[y,\Psi]-\Psi^T\gamma^9[y,\Psi]-
		\Psi^T\gamma^k[x^k,\Psi]\,\right],\label{freef}\\
 \calL^F_{1/2}&=&tr\left[-2\Psi^T[\psi,A]+2\Psi^T\gamma^9[\psi,Y]
		+2\Psi^T\gamma^k[\psi,X^k]\,\right],\label{1f}\\
 \calL^F_{1}&=&tr\Big[\,i\Psi^T\ptau \Psi
		-i\Psi^T\gamma^9\pth \Psi\nn
  &&\hspace{10ex}+\Psi^T[A,\Psi]-\Psi^T\gamma^9[Y,\Psi]
		-\Psi^T\gamma^k[X^k,\Psi]\,\Big].\label{2f}
\end{eqnarray}
Our purpose is to study the behavior of matrix string theory for
small radius $L$. Actually, by using the above action, we perform the
perturbative expansion with respect to $L$ and integrate only the
off-diagonal matrix elements.
The expansion is essentially the strong coupling expansion with
respect to the gauge coupling $g\sim 1/L$.
In the expansion, we regard eqs.(\ref{freeb}) and (\ref{freef}) as the
free parts and eqs.(\ref{1b}), (\ref{2b}), (\ref{1f}) and (\ref{2f})
as the interactions.
In the next subsection, we read off the propagators from the free
parts (\ref{freeb}) and (\ref{freef}).
In subsection \ref{sec:EA}, based on the expansion,
we integrate the off-diagonal matrix elements in eq.(\ref{PI})
and derive the effective action for the diagonal matrix elements,
\begin{eqnarray}
  T&=&\int \calD y \calD x^k \calD\psi \calD c \calD\bar{c}\,\,
	\exp\left(iS_{eff}[y,\,x^k,\,\bar{c},\,c,\,\psi]\right),\\
  &&S_{eff}[y,\,x^k,\,\bar{c},\,c,\,\psi]=LT \int d\tau \int_0^{2\pi}
	d\theta\, \left(\calL^{string}
	-i\ln Z[y,\,x^k,\,\psi]\right),\label{eff}\\
  &&Z[y,\,x^k,\,\psi]=\int \calD A \calD Y \calD X^k \calD\Psi\calD C
	\calD\bar{C}\calD B\,\exp\left[i\tilde{S}\right],\label{Z}\\
  &&\tilde{S}=LT \int d\tau \int_0^{2\pi} d\theta\,
	\left(\calL^B_0 +L\calL^B_1 + L^2\calL^B_2
	 +\calL^F_0 +L^{1/2}\calL^F_{1/2} +L\calL^F_1\right).
\end{eqnarray}
Henceforth we set $\xi=(\tau,\theta)$ and $LT=1$ for brevity.
\subsection{Free action and propagators}
{}From eqs.(\ref{freeb}) and (\ref{freef}), the free action is given by
\begin{eqnarray}
  \tilde{S_0}&=&\int d^2\xi \, (\calL^B_0 +\calL^F_0)\nn\label{free}
  &=&\sum_{a,\,b=1}^N\int d^2\xi\, \Bigg[\,
	   \frac{1}{2}(x_a^k-x_b^k)^2 A_{ab}A_{ba}
	  -\frac{1}{2}(x_a^k-x_b^k)^2 Y_{ab}Y_{ba}
	  -\frac{1}{2}(x_a^k-x_b^k)^2 X_{ab}^l X_{ba}^l\nn
  &&\hspace{30mm}-\frac{1}{2}\left\{(y_a-y_b)Y_{ab}
	+(x_a^k-x_b^k)X_{ab}^k -(y_a-y_b)A_{ab}\right\}\nn
  &&\hspace{25ex}\times\left\{(y_b-y_a)Y_{ba}+(x_b^k-x_a^k)X_{ba}^k
		-(y_b-y_a)A_{ba}\right\}\nn
  &&\hspace{30mm}-iB_{ab}\left\{(y_b-y_a)Y_{ba}
	+(x_b^k-x_a^k)X_{ba}^k-(y_b-y_a)A_{ba}\right\}\nn
  &&\hspace{30mm} -i(x_a^k-x_b^k)^2\bar{C}_{ab}C_{ba}\nn
  &&\hspace{30mm} -(y_{a}-y_b)\Psi_{ab}^T(1-\gamma^9)\Psi_{ba}
		 +(x_{a}^k-x_b^k)\Psi_{ab}^T\gamma^k\Psi_{ba}\Bigg],
\end{eqnarray}
where use has been made of the matrix elements in eqs.(\ref{diag2})
and (\ref{off-diag}). Similarly we can rewrite the interaction
parts (\ref{1b}), (\ref{2b}), (\ref{1f}) and (\ref{2f}) (See appendix
A for the concrete expressions).
{}From the above expression of the free action, it is easy to read off
the propagators,
\begin{eqnarray}
  \vev{Y_{ab}(\xi)Y_{ba}(\xi')} &=&\frac{-i}{(x_a-x_b)^2}
	\left(1-\frac{(y_a-y_b)^2}{(x_a-x_b)^2}\right)
	G(\xi,\xi'),\label{YY}\\
  \vev{X_{ab}^k(\xi)X_{ba}^l(\xi')}&=&\frac{-i}{(x_a-x_b)^2}
	\left(\delta^{kl}-\frac{(x_a^k-x_b^k)(x_a^l-x_b^l)}
	{(x_a-x_b)^2}\right)\,G(\xi,\xi'),\\
  \vev{A_{ab}(\xi)A_{ba}(\xi')}&=&\frac{i}{(x_a-x_b)^2}
	\left(1+\frac{(y_a-y_b)^2}{(x_a-x_b)^2}\right)\,G(\xi,\xi'),\\
  \vev{X_{ab}^k(\xi)Y_{ba}(\xi')} &=&\vev{X_{ab}^k(\xi)A_{ba}(\xi')}
      =i\frac{(x_a^k-x_b^k)(y_a-y_b)}{[(x_a-x_b)^2]^2}\,G(\xi,\xi'),\\
  \vev{A_{ab}(\xi)Y_{ba}(\xi')}&=&
	i\frac{(y_a-y_b)^2}{[(x_a-x_b)^2]^2}\,G(\xi,\xi'),\label{AY}\\
  \vev{B_{ab}(\xi)Y_{ba}(\xi')} &=&\vev{B_{ab}(\xi)A_{ba}(\xi')} =
	\frac{y_a-y_b}{(x_a-x_b)^2}\,G(\xi,\xi'),\\
  \vev{B_{ab}(\xi)X_{ba}^k(\xi')}&=&
	\frac{x_a^k-x_b^k}{(x_a-x_b)^2}\,G(\xi,\xi'),\\
  \vev{\bar{C}_{ab}(\xi)C_{ba}(\xi')}&=&
	\frac{1}{(x_a-x_b)^2}\,G(\xi,\xi'),\\
  \vev{\Psi_{ab}^{\alpha}(\xi)\Psi_{ba}^{\beta}(\xi')}&=& -\frac{i}{2}
	\left\{(y_a-y_b)(I+\gamma^9)_{\alpha\beta}
	+(x_a^k-x_b^k)\gamma^k_{\alpha\beta}\right\}\nn
  &&\hspace{10ex}\times
	\frac{1}{(x_a-x_b)^2}\,G(\xi,\xi')\label{psipsi},
\end{eqnarray}
where
$G(\xi,\xi')\equiv
\delta^{(2)}(\xi-\xi')=\delta(\tau-\tau')\delta(\theta-\theta')$,
$(x_a-x_b)^2\equiv(x_{a}^k-x_b^k)(x_{a}^k-x_b^k)=(x_{a}^k-x_b^k)^2$
and the spinor indices $\alpha,\beta$ run through $1,2, \cdots,16$.
Here we should notice that $(x_{a}-x_b)^2$ for any pairs $a\ne b$
\,($1\leq a,b\leq N$) must be non-zero in order that the perturbative
expansion with respect to $L$ makes sense since the propagators are
singular at $(x_a-x_b)^2=0$.
We recall that in matrix string theory, the usual string interactions
are described by the exchanges of coincident diagonal matrix elements,
which correspond to the world-sheet instanton effects \cite{GHV}.
Hence, at the points where usual string interactions occur,
the perturbative expansion with respect to $L$ does not make sense
even for small radius $L$.
In this paper we consider only the situations in which the
perturbative expansion with respect to $L$ makes sense, and by using
the expansion we integrate out the off-diagonal matrix elements and
derive the effective action for the diagonal matrix elements.
Under these circumstances where the usual string interactions are
neglected, it is expected that the quantum corrections cancel out and
the effective action agrees with the classical (free) action of DLCQ
type IIA superstring.

For later convenience, we define new hatted variables $\hat{X}^K$ and
$\hat{x}^K$ $(K=k,9,10\,\ (k = 1,2,\cdots,8))$,
\begin{equation}
  \hat{X}^k=X^k,\quad\hat{X}^9=Y,\quad\hat{X}^{10}=iA,\quad
  \hat{x}^k=x^k,\quad\hat{x}^9=y,\quad\hat{x}^{10}=iy.\label{hatX}
\end{equation}
The propagators (\ref{YY})-(\ref{AY}) are represented in a single form
with the new variables,
\begin{equation}
  \vev{\hat{X}_{ab}^K(\xi)\hat{X}_{ba}^L(\xi')}
	=\frac{-i}{(\hat{x}_a-\hat{x}_b)^2}
	\left(\delta^{KL}
	-\frac{(\hat{x}_a^K-\hat{x}_b^K)(\hat{x}_a^L-\hat{x}_b^L)}
	{(\hat{x}_a-\hat{x}_b)^2}\right)G(\xi,\xi').\label{pro}
\end{equation}
Here note that $(\hat{x}_a-\hat{x}_b)^2\equiv(\hat{x}_a^K-\hat{x}_b^K)
(\hat{x}_a^K-\hat{x}_b^K)=(x_a^k-x_b^k)(x_a^k-x_b^k)=(x_a-x_b)^2$.

\subsection{Effective action\label{sec:EA}}
\begin{figure}[htbp]
\centerline{\epsfbox{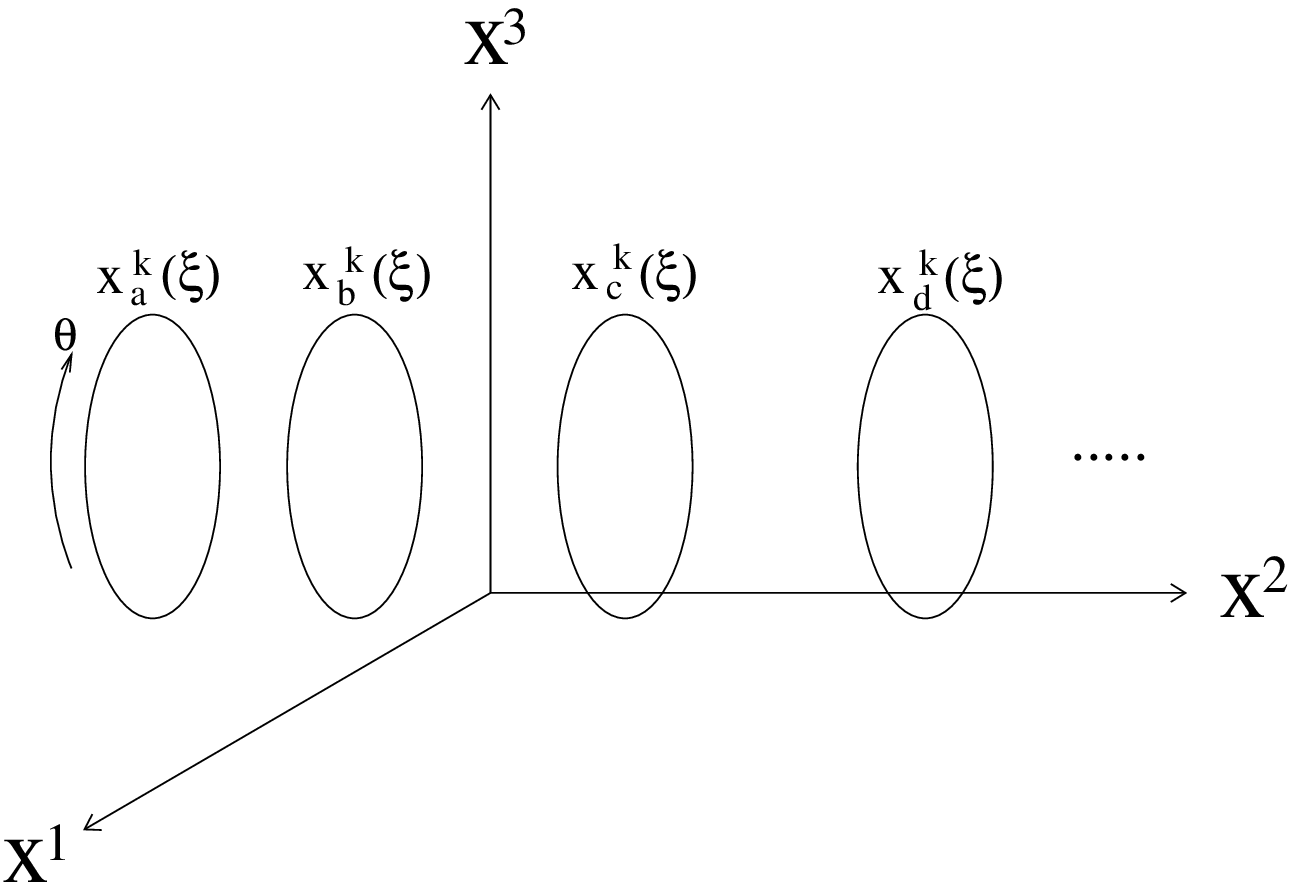}}
\caption{A configuration of $N$ string bits in which the differences
of arbitrary two bits $(x_a^k-x_b^k)$ are constant, though
$x_a^k(\xi)$ and $x_b^k(\xi)$ depend on $\xi$ in general.}
\label{fig1}
\end{figure}

In this subsection we calculate the effective action (\ref{eff})
(or (\ref{Z})) based on the perturbative expansion with respect to
the radius $L$. As is emphasized in Ref.\cite{SY}, however,
the calculations are not well-defined. The reason is as follows:
In the previous subsection we have seen that the propagators
(\ref{YY})-(\ref{psipsi}) in the perturbative expansion are
proportional to the $\delta$-function
$G(\xi,\xi')=\delta^{(2)}(\xi-\xi')$.
Hence, the loop contributions have the ultraviolet divergences like
$\delta^{(2)}(0)$ and we need a regularization for the divergences.
However, it is very difficult to find a suitable regularization
which respects symmetries (e.g., supersymmetry and gauge symmetry)
and hence we cannot perform well-defined calculations.
Actually, we can easily understand a difficulty in finding the
regularization. If we adopt a certain regularization
(e.g., cutoff regularization for large momenta), the regularized
$\delta$-function $G(\xi,\xi')$ would not satisfy such a property of
$\delta$-function that $f(\xi)G(\xi,\xi')=f(\xi')G(\xi,\xi')$.
Then we shall have an ambiguity how we choose the arguments in the
differences of the diagonal matrix elements
$(x_a^k-x_b^k)$ and $(y_a-y_b)$ which appear in the propagators
(\ref{YY})-(\ref{psipsi}).
For example, we can choose $(x_a^k(\xi)-x_b^k(\xi))$ and
$(y_a(\xi)-y_b(\xi))$,~
$(x_a^k(\xi')-x_b^k(\xi'))$ and $(y_a(\xi')-y_b(\xi'))$, or etc..
To avoid the ambiguity, henceforth we consider only the configurations
of the diagonal matrix elements in which the differences of arbitrary
two elements $(x_a^k-x_b^k)$, $(y_a-y_b)$ and
$(\psi_a-\psi_b)$ are independent of $\xi$,
although $x_a^k$, $x_b^k$, $y_a$, $y_b$, $\psi_a$ and $\psi_b$
themselves depend on $\xi$ in general. (See Fig.\ref{fig1} for such a
configuration.)
Here we should notice that we have not yet found the suitable
regularization, though we have reduced an ambiguity by restricting
configurations of the diagonal matrix elements.
Hence we still cannot give the well-defined calculations perfectly
but only give a formal argument about the quantum corrections by
studying whether the coefficients of $\delta^{(2)}(0)$'s cancel out
between the bosonic and fermionic degrees of freedoms.
 
\subsubsection{$O(L^{0})$ }
The lowest order contribution in eq.(\ref{Z}) is the one-loop
determinant of the free action (\ref{free}).
Actually, the determinant is unity due to the coincidence between
bosonic and fermionic degrees of freedoms.

\subsubsection{$O(L^{1/2})$}
The next contribution in eq.(\ref{Z}) comes from the
interaction part (\ref{SF1/2}) of $O(L^{1/2})$.
However the contribution $\vev{i\tilde{S}_{1/2}^F}$ vanishes because
there is no way to self-contract $i\tilde{S}_{1/2}^F$ as we can see
from eq.(\ref{SF1/2}).

\subsubsection{$O(L^{1})$}
The $O(L^{1})$ contributions in eq.(\ref{Z}) come from
eqs.(\ref{SB1}), (\ref{SF1/2}) and (\ref{SF1}).
Actually, there are tree kinds of contributions,
$\vev{i\tilde{S}_{1}^B}$, $(1/2!) \vev{i\tilde{S}_{1/2}^F
\,i\tilde{S}_{1/2}^F}$ and $\vev{i\tilde{S}_{1}^F}$.

First we consider $\vev{i\tilde{S}_{1}^B}$.
In $\vev{i\tilde{S}_{1}^B}$, the second summation in eq.(\ref{SB1})
does not contribute because there is no way to self-contract that
part. Hence, we get
\begin{eqnarray}
\vev{i\tilde{S}_{1}^B}&=&iL\int d^2 \xi \sum_{a,b=1}^N\langle
	i(y_a-y_b)\ptau Y_{ab}Y_{ba} -i(y_a-y_b)\ptau Y_{ab}A_{ba}
	-i(y_a-y_b)\pth A_{ab}Y_{ba}\nn
  &&\hspace{3ex}+i(y_a-y_b)\pth A_{ab}A_{ba}
	+i(y_a-y_b)\ptau X_{ab}^k X_{ba}^k
	-i(x_a^k-x_b^k)\ptau X_{ab}^k A_{ba}\nn
  &&\hspace{3ex}-i(y_a-y_b)\pth X_{ab}^k X_{ba}^k
	+i(x_a^k-x_b^k)\pth X_{ab}^k Y_{ba}
	+B_{ab}\pth Y_{ba}-B_{ab}\ptau A_{ba}\nn
  &&\hspace{3ex}+2(y_a-y_b)\pth \bar{C}_{ab}C_{ba}
	-2(y_a-y_b)\ptau \bar{C}_{ab}C_{ba}\rangle,
\end{eqnarray}
where we have performed partial integrations due to the assumption
that $(x_a^k-x_b^k)$ and $(y_a-y_b)$ are independent of $\xi$.
Furthermore, we can rewrite the equation as
\begin{eqnarray}
\vev{i\tilde{S}_{1}^B} &=&iL\int d^2 \xi \int d^2 \xi'
	 \sum_{a,b=1}^N\bigg\{i(y_a-y_b)\ptau
	\vev{Y_{ab}(\xi)Y_{ba}(\xi')}
	-i(y_a-y_b)\ptau\vev{Y_{ab}(\xi)A_{ba}(\xi')}\nn
  &&\hspace{10ex}-i(y_a-y_b)\pth\vev{A_{ab}(\xi)Y_{ba}(\xi')}
	+i(y_a-y_b)\pth\vev{A_{ab}(\xi)A_{ba}(\xi')}\nn
  &&\hspace{10ex}+i(y_a-y_b)\ptau\vev{X_{ab}^k(\xi)X_{ba}^k(\xi')}
	-i(x_a^k-x_b^k)\ptau\vev{X_{ab}^k(\xi)A_{ba}(\xi')}\nn
  &&\hspace{10ex}-i(y_a-y_b)\pth\vev{X_{ab}^k(\xi)X_{ba}^k(\xi')}
	+i(x_a^k-x_b^k)\pth\vev{X_{ab}^k(\xi)Y_{ba}(\xi')}\nn
  &&\hspace{10ex}-\pth \vev{B_{ab}(\xi)Y_{ba}(\xi')}
	+\ptau\vev{B_{ab}(\xi)A_{ba}(\xi')}\nn
  &&\hspace{10ex}+2(y_a-y_b)\,\pth
	\vev{\bar{C}_{ab}(\xi)C_{ba}(\xi')}\nn
  &&\hspace{25ex}-2(y_a-y_b)\,\ptau\vev{\bar{C}_{ab}(\xi)C_{ba}(\xi')}\bigg\}
	\,\delta^{(2)}(\xi-\xi').
\end{eqnarray}
In this contribution, from the expression of the propagators
(\ref{YY})-(\ref{psipsi}), we see that the quantity in the braces
$\{\hspace{1ex}\}$ is antisymmetric with respect to the exchange of
the indices $a$ and $b$. Hence, $\vev{i\tilde{S}_{1}^B}$ is zero by
summing over $a$ and $b$.

Next we consider
$(1/2!)\vev{i\tilde{S}_{1/2}^F\,i\tilde{S}_{1/2}^F}$.
{}From the interaction (\ref{SF1/2}), we get
\begin{eqnarray}
  &&\hspace{-2ex}\frac{1}{2!} \vev{i\tilde{S}_{1/2}^F \,i\tilde{S}_{1/2}^F}\nn
  &&=-2L\int d^2 \xi \int d^2 \xi'\sum_{a,b=1}^N
	\Bigg[(\psi^{\alpha}_a(\xi)-\psi^{\alpha}_b(\xi))
	(\psi^{\alpha'}_a(\xi')-\psi^{\alpha'}_b(\xi'))\nn
  &&\times\bigg\{
	\vev{\Psi^{\alpha}_{ab}(\xi)\Psi_{ba}^{\alpha'}(\xi')}
	\vev{A_{ba}(\xi)A_{ab}(\xi')}
	-\vev{\Psi^{\alpha}_{ab}(\xi)\Psi_{ba}^{\beta'}(\xi')}
	\vev{A_{ba}(\xi)Y_{ab}(\xi')}\,\gamma^9_{\beta'\alpha'}\nn
  &&\hspace{3ex}
	-\vev{\Psi^{\alpha}_{ab}(\xi)\Psi_{ba}^{\beta'}(\xi')}
	\vev{A_{ba}(\xi)X^k_{ab}(\xi')}\,\gamma^k_{\beta'\alpha'}
	-\vev{\Psi^{\beta}_{ab}(\xi)\Psi_{ba}^{\alpha'}(\xi')}
	\vev{Y_{ba}(\xi)A_{ab}(\xi')}\,\gamma^9_{\beta\alpha}\nn
  &&\hspace{3ex}
	+\vev{\Psi^{\beta}_{ab}(\xi)\Psi_{ba}^{\beta'}(\xi')}
	\vev{Y_{ba}(\xi)Y_{ab}(\xi')}\,\gamma^9_{\beta\alpha}
	\gamma^9_{\beta'\alpha'}
	  +\vev{\Psi^{\beta}_{ab}(\xi)\Psi_{ba}^{\beta'}(\xi')}
	\vev{Y_{ba}(\xi)X^{k'}_{ab}(\xi')}\,\gamma^9_{\beta\alpha}
	\gamma^{k'}_{\beta'\alpha'}\nn
  &&\hspace{3ex}
	-\vev{\Psi^{\beta}_{ab}(\xi) \Psi_{ba}^{\alpha'}(\xi')}
	\vev{X^k_{ba}(\xi)A_{ab}(\xi')}\,\gamma^k_{\beta\alpha}
	+\vev{\Psi^{\beta}_{ab}(\xi)\Psi_{ba}^{\beta'}(\xi')}
	\vev{X^k_{ba}(\xi)Y_{ab}(\xi')}\,\gamma^k_{\beta\alpha}
	\gamma^9_{\beta'\alpha'}\nn
  &&\hspace{3ex}
	+\vev{\Psi^{\beta}_{ab}(\xi)\Psi_{ba}^{\beta'}(\xi')}
	\vev{X^k_{ba}(\xi)X^{k'}_{ab}(\xi')}\,
	\gamma^k_{\beta\alpha}
	\gamma^{k'}_{\beta'\alpha'}\bigg\}\Bigg].
\end{eqnarray}
Also in this contribution,
from the expression of the propagators (\ref{YY})-(\ref{psipsi}),
we see that the quantity in the bracket $[\hspace{2ex}]$
is antisymmetric with respect to the exchange of the indices $a$ and $b$.
Hence, it is obvious that $(1/2!) \vev{i\tilde{S}_{1/2}^F
\,i\tilde{S}_{1/2}^F}$ is zero.

Finally, we consider $\vev{i\tilde{S}_{1}^F}$.
In $\vev{i\tilde{S}_{1}^F}$, the second summation in eq.(\ref{SF1})
does not contribute because there is no way to self-contract that
part. Then we get
\begin{eqnarray}
  \vev{i\tilde{S}_{1}^F}&=& iL\int d^2 \xi \sum_{a,b=1}^N
	\bigg\{i\vev{\Psi_{ab}^T\ptau \Psi_{ba}}
	-i\vev{\Psi_{ab}^T\gamma^9 \pth\Psi_{ba}}\bigg\}\nn
  &=& iL\int d^2 \xi \int d^2 \xi' \sum_{a,b=1}^N
	\bigg\{i\ptau\vev{\Psi_{ab}^{\alpha}(\xi)
			\Psi_{ba}^{\alpha}(\xi')}\nn
  &&\hspace{22ex}-i\pth\vev{\Psi_{ab}^{\alpha}(\xi)
	\Psi_{ba}^{\alpha'}(\xi')}\,\gamma^9_{\alpha\alpha'}
	\bigg\}\delta^{2}(\xi-\xi').
\end{eqnarray}
Also in this contribution, from the expression of the propagators
(\ref{YY})-(\ref{psipsi}), we see that the quantity in the braces
$\{\hspace{1ex}\}$ is antisymmetric with respect to the exchange of
the indices $a$ and $b$. Hence, it is obvious that
$\vev{i\tilde{S}_{1}^F}$ is zero.
Thus, all quantum corrections of $O(L)$ to the classical string action
are zero.
Note that to show the zero quantum corrections of $O(L)$, we have used
only the antisymmetry under the exchange of the indices $a$ and $b$,
and we have never used the fact that $G(\xi,\xi')$, which appears in
the propagators, is the $\delta$-function. Hence the quantum
corrections of $O(L)$ would be zero even if we adopt a certain
regularization and $G(\xi,\xi')$ is the regularized
$\delta$-function.

\subsubsection{$O(L^{3/2})$}
At $O(L^{3/2})$, there are tree kinds of contributions in
eq.(\ref{Z}).
Those are $(1/2!)\vev{i\tilde{S}_{1}^B\,i\tilde{S}_{1/2}^F}$,
$(1/3!)\vev{i\tilde{S}_{1/2}^F\,i\tilde{S}_{1/2}^F\,
i\tilde{S}_{1/2}^F}$ and
$(1/2!)\vev{i\tilde{S}_{1}^F\,i\tilde{S}_{1/2}^F}$.
However, each contribution is zero because there is no way of
contraction in $i\tilde{S}_{1}^B\,i\tilde{S}_{1/2}^F$,
$i\tilde{S}_{1/2}^F\,i\tilde{S}_{1/2}^F\,i\tilde{S}_{1/2}^F$
and $i\tilde{S}_{1}^F\,i\tilde{S}_{1/2}^F$, respectively.

\subsubsection{$O(L^{2})$}
At $O(L^2)$, there are many contributions in eq.(\ref{Z}).
Actually, those are $(1/2!)\vev{i\tilde{S}_{1}^B  i\tilde{S}_{1}^B}$,
$\vev{i\tilde{S}_{2}^B}$,~
$(1/2!)\vev{i\tilde{S}_{1}^F \,i\tilde{S}_{1}^F}$,~
$(1/2!)\vev{i\tilde{S}_1^B i\tilde{S}_1^F}$,~
$(1/3!)\vev{i\tilde{S}_{1/2}^F i\tilde{S}_{1/2}^F i\tilde{S}_{1}^B}$,
$(1/3!)\vev{i\tilde{S}_{1/2}^F i\tilde{S}_{1/2}^F i\tilde{S}_{1}^F}$
and $(1/4!)\vev{ i\tilde{S}_{1/2}^F  i\tilde{S}_{1/2}^F
i\tilde{S}_{1/2}^F  i\tilde{S}_{1/2}^F}$.
Note that the last three contributions contain fermionic diagonal
elements $\psi_a$. They each vanish due to the anti-commutativity of
the Grassmann variable $\psi_a$. And it is easy to show that the
contribution of $(1/2!)\vev{i\tilde{S}_1^B i\tilde{S}_1^F}$ vanishes.
Then $(1/2!)\vev{i\tilde{S}_{1}^B  i\tilde{S}_{1}^B}$,
$\vev{i\tilde{S}_{2}^B}$ and
$(1/2!)\vev{i\tilde{S}_{1}^F \,i\tilde{S}_{1}^F}$
give non-trivial contributions in eq.(\ref{Z}), which will be
calculated below.

\begin{figure}[htbp]
\centerline{\epsfbox{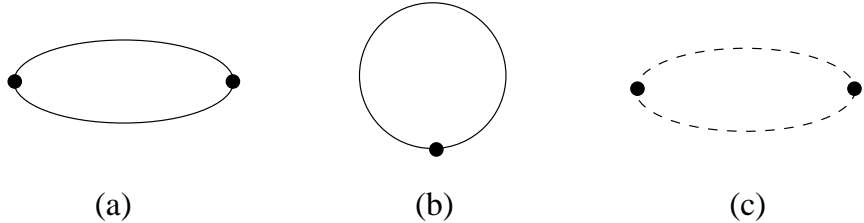}}
\caption{One-loop Feynman diagrams at $O(L^2)$. Figs.(a), (b) and (c)
correspond to one-loop contributions in $(1/2!)\vev{i\tilde{S}_{1}^B
i\tilde{S}_{1}^B}$, $\vev{i\tilde{S}_{2}^B}$ and
$(1/2!)\vev{i\tilde{S}_{1}^F \,i\tilde{S}_{1}^F}$, respectively.
Solid lines denote the propagators of the bosonic or the ghost fields
and dashed lines denote those of the fermionic fields.}
\label{fig2}
\end{figure}

\vspace{\baselineskip}
\noindent (1) One-loop contributions\\
First, we consider the one-loop contributions in
$(1/2!)\vev{i\tilde{S}_{1}^B  i\tilde{S}_{1}^B}$,
$\vev{i\tilde{S}_{2}^B}$ and
$(1/2!)\vev{i\tilde{S}_{1}^F \,i\tilde{S}_{1}^F}$,
which are referred to as
$(1/2!)\vev{i\tilde{S}_{1}^B  i\tilde{S}_{1}^B}^{(one-loop)}$,
$\vev{i\tilde{S}_{2}^B}^{(one-loop)}$ and
$(1/2!)\vev{i\tilde{S}_{1}^F \,i\tilde{S}_{1}^F}^{(one-loop)}$,
respectively.
In Fig.\ref{fig2}, we give Feynman diagrams which correspond to such
contributions.

First of all we consider
$(1/2!)\vev{i\tilde{S}_{1}^B  i\tilde{S}_{1}^B}^{(one-loop)}$, which
is calculated as
\begin{eqnarray}
  &&\hspace{-3ex}\frac{1}{2!}\,\vev{i\tilde{S}_{1}^B
	i\tilde{S}_{1}^B}^{(one-loop)}\nn
  &&=-\frac{L^2}{2}\int d^2 \xi \int d^2 \xi'\sum_{a,b=1}^N
	\sum_{a',b'=1}^N
	\langle\,\Big\{i(y_a-y_b)\ptau Y_{ab}Y_{ba}(\xi)
	-i(y_a-y_b)\ptau Y_{ab}A_{ba}(\xi)\nn
  &&\hspace{20ex}
	-i(y_a-y_b)\pth A_{ab} Y_{ba}(\xi)
	+i(y_a-y_b)\pth A_{ab} A_{ba}(\xi)\nn
  &&\hspace{20ex}
	+i(y_a-y_b)\ptau X_{ab}^k X_{ba}^k(\xi)
	-i(x_a^k-x_b^k)\ptau X_{ab}^k A_{ba}(\xi)\nn
  &&\hspace{20ex}
        -i(y_a-y_b)\pth X_{ab}^k X_{ba}^k(\xi)
	+i(x_a^k-x_b^k)\pth X_{ab}^k Y_{ba}(\xi)\nn
  &&\hspace{20ex}
	+B_{ab}\pth Y_{ba}(\xi) - B_{ab}\ptau A_{ba}(\xi)\nn
  &&\hspace{20ex}
	+2(y_a-y_b)\pth \bar{C}_{ab} C_{ba}(\xi)
	-2(y_a-y_b)\ptau \bar{C}_{ab} C_{ba}(\xi)\Big\}\nn
  &&\hspace{15ex}\times \Big\{i(y_{a'}-y_{b'})
	\ptaup Y_{a'b'}Y_{b'a'}(\xi')
	-i(y_{a'}-y_{b'}) \ptaup Y_{a'b'} A_{b'a'}(\xi')\nn
  &&\hspace{18ex}
	-i(y_{a'}-y_{b'}) \pthp A_{a'b'} Y_{b'a'}(\xi')
	+i(y_{a'}-y_{b'}) \pthp A_{a'b'} A_{b'a'}(\xi')\nn
  &&\hspace{18ex}
	+i(y_{a'}-y_{b'}) \ptaup X_{a'b'}^{k'}
	X_{b'a'}^{k'}(\xi')
	-i(x_{a'}^{k'}-x_{b'}^{k'})
	\ptaup X_{a'b'}^{k'} A_{b'a'}(\xi')\nn
  &&\hspace{18ex}
        -i(y_{a'}-y_{b'})
	\pthp X_{a'b'}^{k'} X_{b'a'}^{k'}(\xi')
	+i(x_{a'}^{k'}-x_{b'}^{k'})
	\pthp X_{a'b'}^{k'} Y_{b'a'}(\xi')\nn
  &&\hspace{18ex}
	+B_{a'b'}\pthp Y_{b'a'}(\xi')-B_{a'b'}
	\ptaup A_{b'a'}(\xi')\nn
  &&\hspace{18ex}
	+2(y_{a'}-y_{b'})\pthp \bar{C}_{a'b'} C_{b'a'}(\xi')
	-2(y_{a'}-y_{b'})\ptaup \bar{C}_{a'b'} C_{b'a'}(\xi')
	\Big\}\,\rangle\nn
  &&=L^2\int d^2 \xi \int d^2 \xi'\sum_{a \ne b} \left[\left(
	\frac{1}{(x_a-x_b)^2}+\frac{17}{2}\,\frac{(y_a-y_b)^2}
	{\{(x_a-x_b)^2\}^2}\right)
	\ptau\ptaup G(\xi,\xi')\,G(\xi,\xi')\right.\nn
  &&\hspace{16ex}
	-17\,\frac{(y_a-y_b)^2}{\{(x_a-x_b)^2\}^2}
	\ptau\pthp G(\xi,\xi')\,G(\xi,\xi')\nn
  &&\hspace{16ex}\left.-\left(
	\frac{1}{(x_a-x_b)^2}-\frac{17}{2}\,
	\frac{(y_a-y_b)^2}{\{(x_a-x_b)^2\}^2}\right)
	\pth\pthp G(\xi,\xi')\,G(\xi,\xi')\right].\label{1-loop1}
\end{eqnarray}

Next we consider $\vev{i\tilde{S}_{2}^B}^{(one-loop)}$ and it is
also calculated as
\begin{eqnarray}
  &&\hspace{-2ex}\vev{i\tilde{S}_{2}^B}^{(one-loop)}\nn
  && = iL^2\int d^2 \xi \sum_{a,b=1}^N\bigg\{
	\frac{1}{2}\vev{\ptau Y_{ab}\ptau Y_{ba}}
	-\vev{\ptau Y_{ab}\pth A_{ba}}
	+\frac{1}{2}\vev{\pth A_{ab} \pth A_{ba}}
	+\frac{1}{2}\vev{\ptau X_{ab}^k \ptau X_{ba}^k}\nn
  &&\hspace{3cm}
	-\frac{1}{2}\vev{\pth X_{ab}^k\pth X_{ba}^k}
	-i\vev{\pth\bar{C}_{ab}\pth C_{ba}}
	+i\vev{\ptau \bar{C}_{ab}\ptau C_{ba}}\bigg\}\nn
  && = L^2\int d^2 \xi \int d^2 \xi'\sum_{a \ne b} \Bigg[
	\left(\frac{3}{(x_a-x_b)^2}
	-\frac{1}{2}\frac{(y_a-y_b)^2}{\{(x_a-x_b)^2\}^2}\right)
	\ptau \ptaup G(\xi,\xi')\,\delta^{(2)}(\xi-\xi')\nn
  &&\hspace{3.3cm}
	+\frac{(y_a-y_b)^2}{\{(x_a-x_b)^2\}^2}\,
	\ptau \pthp G(\xi,\xi')\,\delta^{(2)}(\xi-\xi')\nn
  &&\hspace{3.3cm}
	-\left(\frac{3}{(x_a-x_b)^2}
	+\frac{1}{2}\frac{(y_a-y_b)^2}{\{(x_a-x_b)^2\}^2}\right)
	\pth\pthp G(\xi,\xi')\,\delta^{(2)}(\xi-\xi')\Bigg].
	\label{1-loop2}
\end{eqnarray}

Finally, we consider $(1/2!) \vev{i\tilde{S}_{1}^F
i\tilde{S}_{1}^F}^{(one-loop)}$. It is given by
\begin{eqnarray}
  &&\hspace{-2ex}\frac{1}{2!}\vev{i\tilde{S}_{1}^F
	i\tilde{S}_{1}^F}^{(one-loop)}\nn
  &&=-\frac{L^2}{2}\int d^2 \xi \int d^2 \xi' \sum_{a,b=1}^N
	\sum_{a',b'=1}^N \langle \{i\Psi_{ab}^T\ptau \Psi_{ba}(\xi)
	-i\Psi_{ab}^T\gamma^9\pth \Psi_{ba}(\xi)\}\nn
  &&\hspace{28ex}\times\{i\Psi_{a'b'}^T\ptaup \Psi_{b'a'}(\xi')
	-i\Psi_{a'b'}^T\gamma^9\pthp \Psi_{b'a'}(\xi')\}\rangle\nn
  && =L^2\int d^2 \xi \int d^2 \xi' \sum_{a \ne b}\Bigg[
	\left(\frac{-4}{(x_a-x_b)^2}
	-8\frac{(y_a-y_b)^2}{\{(x_a-x_b)^2\}^2}\right)
	\ptau \ptaup G(\xi,\xi')\,G(\xi,\xi')\nn
  &&\hspace{3.3cm}+16\,\frac{(y_a-y_b)^2}{\{(x_a-x_b)^2\}^2}\,
	\ptau \pthp G(\xi,\xi')\,G(\xi,\xi')\nn
  &&\hspace{3.3cm}+\left(\frac{4}{(x_a-x_b)^2}
	-8\,\frac{(y_a-y_b)^2}{\{(x_a-x_b)^2\}^2}\right)
	\pth \pthp G(\xi,\xi')\,G(\xi,\xi')\Bigg]\label{1-loop3}.
\end{eqnarray}
Note that we have never used the fact that $G(\xi,\xi')$ is the
$\delta$-function in calculating eqs.(\ref{1-loop1})-(\ref{1-loop3}).
Hence eqs.(\ref{1-loop1})-(\ref{1-loop3}) is expected to be unaltered
even if we adopt a certain regularization and $G(\xi,\xi')$ is a
regularized $\delta$-function.
We first use the fact that $G(\xi,\xi')$ is the $\delta$-function at
this stage and it is shown that the one-loop quantum correction at
$O(L^2)$ is zero, i.e.,
$(1/2!)\vev{i\tilde{S}_{1}^B  i\tilde{S}_{1}^B}^{(one-loop)}+
\vev{i\tilde{S}_{2}^B}^{(one-loop)}+ (1/2!)\vev{i\tilde{S}_{1}^F
\,i\tilde{S}_{1}^F}^{(one-loop)}=0$.
Of course the above calculations in matrix string theory are
essentially the same as the ones in the wrapped supermembrane theory
\cite{SY}\footnote{In Ref.\cite{SY}, the zero-mode gauge field $a$ is
restricted to be zero by hand in calculating the one-loop quantum
corrections. In this paper, however, we have just fixed the gauge ($a=y$)
and added the
corresponding FP-ghost part following the standard procedure
\cite{KU}. In this sense the configuration of the gauge field $a$ is
not restricted in our calculations.}.

\begin{figure}[htbp]
\centerline{\epsfbox{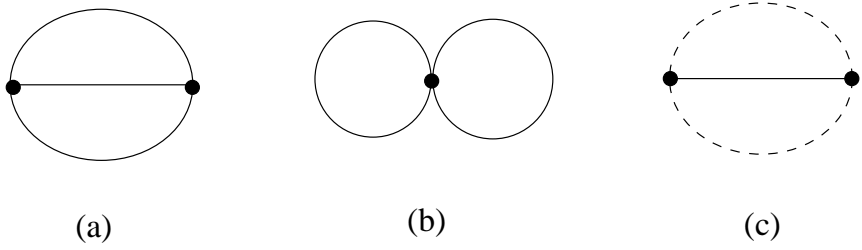}}
\caption{Two-loop Feynman diagrams at $O(L^2)$.
Figs.(a), (b) and (c)  correspond to two-loop contributions in
$(1/2!)\vev{i\tilde{S}_{1}^B  i\tilde{S}_{1}^B}$,
$\vev{i\tilde{S}_{2}^B}$ and
$(1/2!) \vev{i\tilde{S}_{1}^F \,i\tilde{S}_{1}^F}$, respectively.}
\label{fig3}
\end{figure}

\vspace{\baselineskip}
\noindent (2) Two-loop contributions\\
Next, we consider the two-loop contributions in
$(1/2!)\vev{i\tilde{S}_{1}^B  i\tilde{S}_{1}^B}$,
$\vev{i\tilde{S}_{2}^B}$ and $(1/2!)
\vev{i\tilde{S}_{1}^F\,i\tilde{S}_{1}^F}$,
which are not calculated in Ref.\cite{SY}.
We refer to them as
$(1/2!)\vev{i\tilde{S}_{1}^B  i\tilde{S}_{1}^B}^{(two-loop)}$,
$\vev{i\tilde{S}_{2}^B}^{(two-loop)}$ and
$(1/2!) \vev{i\tilde{S}_{1}^F \,i\tilde{S}_{1}^F}^{(two-loop)}$,
respectively. In Fig.\ref{fig3}, we give Feynman diagrams
corresponding to them.

First, we consider
$(1/2!) \vev{i\tilde{S}_{1}^B  i\tilde{S}_{1}^B}^{(two-loop)}$, which
is given by
\begin{eqnarray}
  &&\hspace{-2ex}
    \frac{1}{2!}\vev{i\tilde{S}_1^B  i\tilde{S}_1^B}^{(two-loop)}\nn
  &&=-\frac{L^2}{2}\int d^2 \xi \int d^2 \xi' \sum_{a,b,c=1}^N
	\sum_{a',b',c'=1}^N \langle\,
	\Big\{ -(y_a-y_b) Y_{ab}(A_{bc}Y_{ca}-Y_{bc}A_{ca})(\xi)\nn
  &&\hspace{10ex}
	+(y_a-y_b)A_{ab} (A_{bc}Y_{ca}-Y_{bc}A_{ca})(\xi)
	-(y_a-y_b)X_{ab}^k (A_{bc}X_{ca}^k-X_{bc}^k A_{ca})(\xi)\nn
  &&\hspace{10ex}
	+(x_a^k-x_b^k)A_{ab}(A_{bc}X_{ca}^k-X_{bc}^k A_{ca})(\xi)
	+(y_a-y_b)X_{ab}^k(Y_{bc}X_{ca}^k-X_{bc}^k Y_{ca})(\xi)\nn
  &&\hspace{10ex}
	-(x_a^k-x_b^k)Y_{ab} (Y_{bc}X_{ca}^k-X_{bc}^k Y_{ca})(\xi)
	+(x_a^k-x_b^k)X_{ab}^l (X_{bc}^k X_{ca}^l
	-X_{bc}^l X_{ca}^k)(\xi)\nn
  &&\hspace{10ex}
      -i(y_a-y_b)\bar{C}_{ab} (A_{bc}C_{ca}-C_{bc}A_{ca})(\xi)
      +i(y_a-y_b)\bar{C}_{ab} (Y_{bc}C_{ca}-C_{bc}Y_{ca})(\xi)\nn
  &&\hspace{10ex}
	+i(x_a^k-x_b^k) \bar{C}_{ab} (X_{bc}^k
	C_{ca}-C_{bc}X_{ca}^k)(\xi) \Big\}\nn
  &&\hspace{8ex}\times \Big\{-(y_{a'}-y_{b'})\,Y_{a'b'}
	(A_{b'c'} Y_{c'a'} - Y_{b'c'} A_{c'a'})(\xi')\nn
  &&\hspace{12ex}
	+(y_{a'}-y_{b'})\,A_{a'b'} (A_{b'c'}Y_{c'a'}
	-Y_{b'c'}A_{c'a'})(\xi')\nn
  &&\hspace{12ex}
	-(y_{a'}-y_{b'})\,X_{a'b'}^{k'}
	(A_{b'c'} X_{c'a'}^{k'} - X_{b'c'}^{k'} A_{c'a'})(\xi')\nn
  &&\hspace{12ex}
	+(x_{a'}^{k'}-x_{b'}^{k'})\,A_{a'b'}
	(A_{b'c'} X_{c'a'}^{k'} - X_{b'c'}^{k'} A_{c'a'})(\xi')\nn
  &&\hspace{12ex}
	+(y_{a'}-y_{b'})\,X_{a'b'}^{k'}
	(Y_{b'c'} X_{c'a'}^{k'} - X_{b'c'}^{k'} Y_{c'a'})(\xi')\nn
  &&\hspace{12ex}
	-(x_{a'}^{k'}-x_{b'}^{k'})\,Y_{a'b'}
	(Y_{b'c'} X_{c'a'}^{k'} - X_{b'c'}^{k'} Y_{c'a'})(\xi')\nn
  &&\hspace{12ex}
	+(x_{a'}^{k'}-x_{b'}^{k'})\,X_{a'b'}^{l'}
	(X_{b'c'}^{k'} X_{c'a'}^{l'}
	-X_{b'c'}^{l'} X_{c'a'}^{k'})(\xi')\nn
  &&\hspace{12ex}
      -i(y_{a'}-y_{b'})\,\bar{C}_{a'b'}(A_{b'c'}C_{c'a'}
	-C_{b'c'} A_{c'a'})(\xi')\nn
  &&\hspace{12ex}
	+i(y_{a'}-y_{b'})\,\bar{C}_{a'b'}
	(Y_{b'c'} C_{c'a'}-C_{b'c'} Y_{c'a'})(\xi')\nn
  &&\hspace{12ex}
	+i(x_{a'}^{k'}-x_{b'}^{k'})\,\bar{C}_{a'b'}
	(X_{b'c'}^{k'}C_{c'a'}
	-C_{b'c'} X_{c'a'}^{k'})(\xi')\Big\}\rangle.
\end{eqnarray}
By using the variables $\hat{X}^K$ and $\hat{x}^K$ in eq.(\ref{hatX})
and the propagator (\ref{pro}), we can put the above expression into a
compact form and we obtain
\begin{eqnarray}
  &&\hspace{-2ex}\frac{1}{2!}\vev{i\tilde{S}_{1}^B
	i\tilde{S}_{1}^B}^{(two-loop)}\nn
  &&=-\frac{L^2}{2}\int d^2 \xi \int d^2 \xi'\sum_{a,b,c=1}^N
	\sum_{a',b',c'=1}^N \langle\, \Big\{
	(\hat{x}_a^K-\hat{x}_b^K)\hat{X}_{ab}^L
	(\hat{X}_{bc}^K\hat{X}_{ca}^L
	-\hat{X}_{bc}^L\hat{X}_{ca}^K)(\xi)\nn
  &&\hspace{0.5cm}
	+i(\hat{x}_a^K-\hat{x}_b^K)\bar{C}_{ab}
	(\hat{X}_{bc}^K C_{ca}-C_{bc}\hat{X}_{ca}^K)(\xi)\Big\}\,
	\Big\{(\hat{x}_{a'}^{K'}-\hat{x}_{b'}^{K'})
	\hat{X}_{a'b'}^{L'}(\hat{X}_{b'c'}^{K'}\hat{X}_{c'a'}^{L'}
	-\hat{X}_{b'c'}^{L'}\hat{X}_{c'a'}^{K'})(\xi')\nn
  &&\hspace{30ex}
	+i(\hat{x}_{a'}^{K'}-\hat{x}_{b'}^{K'})\bar{C}_{a'b'}
	(\hat{X}_{b'c'}^{K'}C_{c'a'}
	-C_{b'c'}\hat{X}_{c'a'}^{K'})(\xi')\Big\}\rangle\nn
  &&=-i L^2\int d^2 \xi \int d^2 \xi'\sum_{a\ne b,\,b\ne c,\,c\ne a}
	\left\{\frac{33}{2}\frac{1}{(\hat{x}_a-\hat{x}_b)^2
	(\hat{x}_b-\hat{x}_c)^2}\right.\nn
  &&\hspace{32ex}-16\,\frac{\{(\hat{x}_a^K-\hat{x}_b^K)
	(\hat{x}_c^K-\hat{x}_a^K)\}^2}{(\hat{x}_a-\hat{x}_b)^2
	(\hat{x}_b-\hat{x}_c)^2\{(\hat{x}_c-\hat{x}_a)^2\}^2}\nn
  &&\hspace{32ex}-\frac{1}{2}\,\frac{\{(\hat{x}_a^K-\hat{x}_b^K)
	(\hat{x}_b^K-\hat{x}_c^K)\}^2}{\{(\hat{x}_a-\hat{x}_b)^2
	(\hat{x}_b-\hat{x}_c)^2\}^2}\Bigg\}(G(\xi,\xi'))^3\nn
  &&=-i L^2\int d^2 \xi \int d^2 \xi'\sum_{a\ne b,\,b\ne c,\,c\ne a}
	\Bigg\{\frac{33}{2}\frac{1}{(x_a-x_b)^2(x_b-x_c)^2}\nn
   &&\hspace{32ex}-16\,\frac{\{(x_a^k-x_b^k)(x_c^k-x_a^k)\}^2}
	{(x_a-x_b)^2(x_b-x_c)^2\{(x_c-x_a)^2\}^2}\nn
   &&\hspace{32ex}
	-\frac{1}{2}\,\frac{\{(x_a^k-x_b^k)(x_b^k-x_c^k)\}^2}
	{\{(x_a-x_b)^2(x_b-x_c)^2\}^2}
	\Bigg\}\,(G(\xi,\xi'))^3.\label{1'}
\end{eqnarray}

Next, we consider $\vev{i\tilde{S}_{2}^B}^{(two-loop)}$,
\begin{eqnarray}
  &&\hspace{-2ex}\vev{i\tilde{S}_{2}^B}^{(two-loop)}\nn
  &&=iL^2\int d^2 \xi \Bigg[\sum_{a,b,c,d=1}^N \bigg\{
	-\vev{A_{ab}Y_{bc}A_{cd}Y_{da}}
	+ \vev{A_{ab}Y_{bc}Y_{cd}A_{da}}
       -\vev{A_{ab}X_{bc}^k A_{cd}X_{da}^k}\nn
  &&\hspace{20ex}
       +\vev{A_{ab}X_{bc}^k X_{cd}^k A_{da}}
       +\vev{Y_{ab}X_{bc}^k Y_{cd}X_{da}^k}
       - \vev{Y_{ab}X_{bc}^k X_{cd}^k Y_{da}}\nn
  &&\hspace{20ex}
       +\frac{1}{2}\vev{X_{ab}^k X_{bc}^l X_{cd}^k X_{da}^l}
       -\frac{1}{2}\vev{X_{ab}^k X_{bc}^l X_{cd}^l X_{da}^k}\bigg\}\nn
  &&\hspace{6ex}+\sum_{a,b,c=1}^N\bigg\{
	i\vev{\bar{C}_{ab}A_{ba}C_{ac}A_{ca}}
	-i\vev{\bar{C}_{ab}A_{ba}A_{ac}C_{ca}}
	 -i\vev{A_{ab}\bar{C}_{ba}C_{ac}A_{ca}}\nn
  &&\hspace{14ex}
	 +i\vev{A_{ab}\bar{C}_{ba}A_{ac}C_{ca}}
          -i\vev{\bar{C}_{ab}Y_{ba}C_{ac}Y_{ca}}
	 +i\vev{\bar{C}_{ab}Y_{ba}Y_{ac}C_{ca}}\nn
  &&\hspace{14ex}
	 +i\vev{Y_{ab}\bar{C}_{ba}C_{ac}Y_{ca}}
	 -i\vev{Y_{ab}\bar{C}_{ba}Y_{ac}C_{ca}}
        -i\vev{\bar{C}_{ab}X_{ba}^kC_{ac}X_{ca}^k}\nn
 &&\hspace{14ex}
	 +i\vev{\bar{C}_{ab}X_{ba}^k X_{ac}^kC_{ca}}
	 +i\vev{X_{ab}^k\bar{C}_{ba}C_{ac}X_{ca}^k}
	 -i\vev{X_{ab}^k\bar{C}_{ba}X_{ac}^kC_{ca}}
	\bigg\}\Bigg].
\end{eqnarray}
By using the variables $\hat{X}^K$ and $\hat{x}^K$ in eq.(\ref{hatX})
and the propagator (\ref{pro}), we can also put the above expression
into a compact form and we get
\begin{eqnarray}
  &&\hspace{-2ex}\vev{i\tilde{S}_{2}^B}^{(two-loop)}\nn
  &&=iL^2\int d^2 \xi \Bigg[\sum_{a,b,c,d=1}^N \bigg\{
	\frac{1}{2}\vev{\hat{X}_{ab}^K\hat{X}_{bc}^L
	\hat{X}_{cd}^K\hat{X}_{da}^L}
	-\frac{1}{2}\vev{\hat{X}_{ab}^K\hat{X}_{bc}^L
	\hat{X}_{cd}^L\hat{X}_{da}^K}  \bigg\}\nn
  &&\hspace{12ex} +\sum_{a,b,c=1}^N\bigg\{
        -i\vev{\bar{C}_{ab}\hat{X}_{ba}^KC_{ac}\hat{X}_{ca}^K}
	+i\vev{\bar{C}_{ab}\hat{X}_{ba}^K\hat{X}_{ac}^KC_{ca}}\nn
  &&\hspace{22ex}
	+i\vev{\hat{X}_{ab}^K\bar{C}_{ba}C_{ac}\hat{X}_{ca}^K}
	-i\vev{\hat{X}_{ab}^K\bar{C}_{ba}\hat{X}_{ac}^KC_{ca}}
	\bigg\}\Bigg]\nn
  &&= iL^2\int d^2 \xi \int d^2\xi'\Bigg[\sum_{a\ne b,\, b\ne c}
	\left\{\frac{73}{2}\,\frac{1}{(\hat{x}_a-\hat{x}_b)^2
	(\hat{x}_b-\hat{x}_c)^2}
	-\frac{1}{2}\,\frac{\{(\hat{x}_a^K
	-\hat{x}_b^K)(\hat{x}_b^K-\hat{x}_c^K)\}^2}{\{(\hat{x}_a
	-\hat{x}_b)^2(\hat{x}_b-\hat{x}_c)^2\}^2}\right\}\nn
  &&\hspace{20ex}
	+18\,\sum_{a\ne b} \frac{1}{\{(\hat{x}_a-\hat{x}_b)^2\}^2}
	\Bigg]\,(G(\xi,\xi'))^2\,\delta^{(2)}(\xi-\xi')\nn
  &&=iL^2\int d^2 \xi \int d^2\xi'\Bigg[\sum_{a\ne b,\, b\ne c}
	\left\{\frac{73}{2}\frac{1}{(x_a-x_b)^2(x_b-x_c)^2}
	-\frac{1}{2}\frac{\{(x_a^k-x_b^k)(x_b^k-x_c^k)\}^2}
	{\{(x_a-x_b)^2(x_b-x_c)^2\}^2}\right\}\nn
  &&\hspace{20ex}
	+18 \sum_{a\ne b} \frac{1}{\{(x_a-x_b)^2\}^2}
	\Bigg]\,(G(\xi,\xi'))^2\,\delta^{(2)}(\xi-\xi').\label{eq:2B2}
\end{eqnarray}
The first and second summations in the above equation correspond to
the third and forth summations in eq.(\ref{SB2}), respectively.
Now we extract the $a=c$ part from the first summation in
eq.(\ref{eq:2B2}) and add it to the second summation. Then we get
\begin{eqnarray}
  &&\hspace{-2ex}\vev{i\tilde{S}_{2}^B}^{(two-loop)}\nn
  &&=iL^2\int d^2 \xi \int d^2\xi'\Bigg[
	\sum_{a\ne b,\, b\ne c,\,c\ne a}
	\left\{\frac{73}{2}\frac{1}{(x_a-x_b)^2(x_b-x_c)^2}
	-\frac{1}{2}\frac{\{(x_a^k-x_b^k)(x_b^k-x_c^k)\}^2}
	{\{(x_a-x_b)^2(x_b-x_c)^2\}^2}\right\}\nn
  &&\hspace{20ex}
	+54 \sum_{a\ne b} \frac{1}{\{(x_a-x_b)^2\}^2}
	\Bigg]\,(G(\xi,\xi'))^2\,\delta^{(2)}(\xi-\xi').\label{2'}
\end{eqnarray}

Finally we consider
$(1/2!) \vev{i\tilde{S}_{1}^F \,i\tilde{S}_{1}^F}^{(two-loop)}$
and the contribution is calculated as
\begin{eqnarray}
  &&\hspace{-2ex}\frac{1}{2!} \vev{i\tilde{S}_{1}^F
	\,i\tilde{S}_{1}^F}^{(two-loop)}\nn
  &&=-\frac{L^2}{2}\int d^2 \xi \int d^2 \xi'
	\sum_{a,b,c=1}^N \sum_{a',b',c'=1}^N\langle\,
	\bigg\{\Psi_{ab}^T(A_{bc}\Psi_{ca}-\Psi_{bc}A_{ca})(\xi)\nn
  &&\hspace{15ex}
	-\Psi_{ab}^T\gamma^9(Y_{bc}\Psi_{ca}-\Psi_{bc}Y_{ca})(\xi)
	-\Psi_{ab}^T\gamma^k(X_{bc}^k\Psi_{ca}
	-\Psi_{bc}X_{ca}^k)(\xi) \bigg\}\nn
  &&\hspace{12ex}\times \bigg\{\Psi_{a'b'}^T(A_{b'c'}\Psi_{c'a'}
	-\Psi_{b'c'}A_{c'a'})(\xi')
	-\Psi_{a'b'}^T\gamma^9(Y_{b'c'}\Psi_{c'a'}
	-\Psi_{b'c'}Y_{c'a'})(\xi')\nn
  &&\hspace{20ex}
	-\Psi_{a'b'}^T\gamma^{k'}(X_{b'c'}^{k'}\Psi_{c'a'}
	-\Psi_{b'c'}X_{c'a'}^{k'})(\xi')\bigg\}\rangle\nn
  &&=-iL^2\int d^2 \xi \int d^2 \xi'\sum_{a\ne b,\,b\ne c,\,c\ne a}
	\bigg\{20\,\frac{1}{(x_a-x_b)^2(x_b-x_c)^2}\nn
   &&\hspace{27ex}+16\,\frac{\{(x_a^k-x_b^k)(x_c^k-x_a^k)\}^2}
	{(x_a-x_b)^2(x_b-x_c)^2\{(x_c-x_a)^2\}^2}
	\bigg\}(G(\xi,\xi'))^3. \label{3'}
\end{eqnarray}
Note that in calculating eqs.(\ref{1'}), (\ref{2'}) and 
(\ref{3'}) we have never used
the fact that $G(\xi,\xi')$ is the $\delta$-function.
Hence eqs.(\ref{1'}), (\ref{2'}) and (\ref{3'}) are expected to be unaltered
even if we adopt a certain regularization and $G(\xi,\xi')$
is a regularized $\delta$-function.
We first use the fact that $G(\xi,\xi')$ is the $\delta$-function at this
stage and sum up eqs.(\ref{1'}), (\ref{2'}) and (\ref{3'}). Then, we obtain
\begin{eqnarray}
  &&\frac{1}{2!}\vev{i\tilde{S}_{1}^B
	  i\tilde{S}_{1}^B}^{(two-loop)}+
	\vev{i\tilde{S}_{2}^B}^{(two-loop)}+
	\frac{1}{2!} \vev{i\tilde{S}_{1}^F
	  \,i\tilde{S}_{1}^F}^{(two-loop)}\nn
 &&\hspace{2ex}=iL^2\int d^2 \xi \int d^2\xi'\sum_{a\ne b}
	\frac{54}{\{(x_a-x_b)^2\}^2}\,(G(\xi,\xi'))^3.\label{eq:2Lsum}
\end{eqnarray}
Thus we see that the two-loop quantum corrections at $O(L^2)$ do not
cancel out. One comment is in order:  The remaining term is exactly
that of the second summation in eq.(\ref{2'}).
If we assume that the differences of the diagonal elements can be
estimated as $(x_a^k-x_b^k)\sim O(N^{\alpha})$\footnote{According to
the correspondence of a long string in matrix string theory
with the wrapped supermembrane given in Ref.\cite{SY},
$\alpha=-1$ for $|a-b|\ll N$.} with some
common constant $\alpha$ when $N$ is large, we will see that the terms
canceled in eq.(\ref{eq:2Lsum}), i.e., terms given by the summations
over $a,b$ and $c$ with $a\ne b$, $b\ne c$, $c\ne a$ in
eqs.(\ref{1'}), (\ref{2'}) and (\ref{3'}), behave as
$\sum_{a\ne b,\,b\ne c,\,c\ne a} (x_a^k-x_b^k)^{-2}
(x_b^k-x_c^k)^{-2} \sim O(N^{3-4\alpha})$, while
the remaining term, which comes from the second summation in
eq.(\ref{2'}), behaves as $\sum_{a\ne b} (x_a^k-x_b^k)^{-4}\sim
O(N^{2-4\alpha})$.
In this sense, we could say that only the leading terms in the large
$N$ can be canceled out in the two-loop quantum corrections to the
classical string action at $O(L^2)$.

It will be pedagogical to re-consider the results
(\ref{1'}), (\ref{2'}) and (\ref{3'}) in the case of $N=2$.
In this case, it is obvious that eqs.(\ref{1'}) and (\ref{3'})
are zero. The reason is as follows:
Schematically, each term in eqs.(\ref{1'}) and (\ref{3'}) is
represented by $\vev{tr(X^3) tr(X^3)}$, where $X$ stands for a bosonic
or fermionic $2\times 2$ matrix of only off-diagonal components, i.e.,
its diagonal components are zero. Thus $tr(X^3)=0$ and hence
eqs.(\ref{1'}) and (\ref{3'}) are zero. On the other hand, each term
in eq.(\ref{2'}) is schematically represented by $\vev{tr(X^4)}$ and
it can have non-zero value.
Thus in $N=2$ case, it is obvious that only the bosonic
contribution of eq.(\ref{2'}) exists.

\section{Conclusion and discussion\label{sec:CandD}}
In this paper we have studied in matrix string theory whether the
reduction  to the diagonal elements of the matrices is justified
quantum mechanically. We have seen that at $O(L^2)$, the two-loop
quantum corrections do not cancel out.
Our calculations are essentially two-loop extension of the previous
ones in Ref.\cite{SY}.
 
We should note that no suitable regularization for the divergences
of $\delta^{(2)}(0)$ type is found so far, and hence we have only
studied a mechanism of cancellations of the divergences between
bosonic and fermionic degrees of freedoms.
Actually, we have found that at the two-loop level of $O(L^2)$,
the sub-leading term in the large $N$ appears only from the bosonic
degrees of freedom and cannot be canceled out.
Even if we find a suitable regularization, such a structure seems to
be unaltered and hence our result will be unchanged.

Finally, we comment on the global constraints (\ref{global1}) and
(\ref{global2}) in the wrapped supermembrane theory.
To be precise, such constraints should be taken into account in the
calculations of the quantum double-dimensional reduction\footnote{
In Ref.\cite{SY}, the global constraints are not considered in their
calculations.}.
In matrix string theory, however, there are no counterparts of such
constraints, as was discussed in section \ref{sec:MtoS}.
In particular, in the standard derivation of matrix string theory,
they do not appear naturally.
However, our result may suggest that the suitably matrix-regularized
constraints should be incorporated with the standard form of matrix
string theory.\\

\noindent {\bf Acknowledgments:}
The work of SU is supported in part by the Grant-in-Aid for Scientific
Research No.13135212.

\appendix
\section{Interaction part of the action}
In this appendix we give the interaction part of the
action by using the matrix elements in eqs.(\ref{diag2}) and
(\ref{off-diag}),
\begin{eqnarray}
  \tilde{S}_1^B&=&L\int d^2 \xi\, \calL_1^B\nn
  &=&L\int d^2 \xi \Bigg[\sum_{a,b=1}^N\bigg\{
	i(y_a-y_b)\ptau Y_{ab}Y_{ba}-2i(y_a-y_b)\ptau Y_{ab}A_{ba}
	+i(y_a-y_b)\ptau A_{ab}Y_{ba}\nn
  &&\hspace{2cm}
	-2i(y_a-y_b)\pth A_{ab}Y_{ba}+i(y_a-y_b)\pth A_{ab}A_{ba}
	+i(y_a-y_b)\pth Y_{ab}A_{ba}\nn
  &&\hspace{2cm}
	+i(y_a-y_b)\ptau X_{ab}^k X_{ba}^k
	-2i(x_a^k-x_b^k)\ptau X_{ab}^k A_{ba}
	+i(x_a^k-x_b^k)\ptau A_{ab}X_{ba}^k\nn
  &&\hspace{2cm}
	-i(y_a-y_b)\pth X_{ab}^k X_{ba}^k
	+2i(x_a^k-x_b^k)\pth X_{ab}^k Y_{ba}
	-i(x_a^k-x_b^k)\pth Y_{ab}X^k_{ba}\nn
  &&\hspace{2cm}
	+B_{ab}\pth Y_{ba}-B_{ab}\ptau A_{ba}
	+(y_a-y_b)\pth \bar{C}_{ab}C_{ba}
	-(y_a-y_b)\bar{C}_{ab}\pth C_{ba}\nn
  &&\hspace{2cm}
	+(y_a-y_b)\bar{C}_{ab}\ptau C_{ba}
	-(y_a-y_b)\ptau \bar{C}_{ab}C_{ba}\bigg\}\nn
  &&\hspace{3ex}
	+\sum_{a,b,c=1}^N\bigg\{
	-(y_a-y_b)Y_{ab}(A_{bc}Y_{ca}-Y_{bc}A_{ca})
	+(y_a-y_b)A_{ab}(A_{bc}Y_{ca}-Y_{bc}A_{ca})\nn
  &&\hspace{13ex}
	- (y_a-y_b)X_{ab}^k(A_{bc}X_{ca}^k-X_{bc}^k A_{ca})
	+(x_a^k-x_b^k)A_{ab}(A_{bc}X_{ca}^k-X_{bc}^k A_{ca})\nn
  &&\hspace{13ex}
	+(y_a-y_b)X_{ab}^k(Y_{bc}X_{ca}^k-X_{bc}^k Y_{ca})
	-(x_a^k-x_b^k)Y_{ab}(Y_{bc}X_{ca}^k-X_{bc}^k Y_{ca})\nn
  &&\hspace{13ex}
	+(x_a^k-x_b^k)X_{ab}^l(X_{bc}^k X_{ca}^l-X_{bc}^l X_{ca}^k)
	-i(y_a-y_b)\bar{C}_{ab}(A_{bc}C_{ca}-C_{bc}A_{ca})\nn
  &&\hspace{13ex}
	+i(y_a-y_b)\bar{C}_{ab}(Y_{bc}C_{ca}-C_{bc}Y_{ca})\nn
  &&\hspace{13ex}
	+i(x_a^k-x_b^k)\bar{C}_{ab}(X_{bc}^kC_{ca}-C_{bc}X_{ca}^k)
	\bigg\}\Bigg],\label{SB1}\\
  \tilde{S}_2^B&=&L^2\int d^2 \xi\, \calL_2^B\nn
  &=& L^2\int d^2 \xi \left[\sum_{a,b=1}^N\bigg\{
	\frac{1}{2}\ptau Y_{ab}\ptau Y_{ba}
	-\ptau Y_{ab}\pth A_{ba}
	+\frac{1}{2}\pth A_{ab}\pth A_{ba}
	+\frac{1}{2}\ptau X_{ab}^k\ptau X_{ba}^k\right.\nn
  &&\hspace{3cm}
          -\frac{1}{2}\pth X_{ab}^k\pth X_{ba}^k
	  -i\pth \bar{C}_{ab}\pth C_{ba}
	  +i\ptau \bar{C}_{ab}\ptau C_{ba}\bigg\}\nn
  &&\hspace{1cm}
        +\sum_{a,b,c=1}^N\bigg\{
	i\pth A_{ab}(A_{bc}Y_{ca}-Y_{bc}A_{ca})
	-i\ptau Y_{ab}(A_{bc}Y_{ca}-Y_{bc}A_{ca})\nn
  &&\hspace{15ex}
       -i\ptau X_{ab}^k(A_{bc}X_{ca}^k-X_{bc}^k A_{ca})
       +i\pth X_{ab}^k(Y_{bc}X_{ca}^k-X_{bc}^k Y_{ca})\nn
  &&\hspace{15ex}
       -\pth\bar{C}_{ab}(Y_{bc}C_{ca}-C_{bc}Y_{ca})
       +\ptau\bar{C}_{ab}(A_{bc}C_{ca}-C_{bc}A_{ca})\bigg\}\nn
  &&\hspace{1cm}
       +\sum_{a,b,c,d=1}^N\bigg\{
       -A_{ab}Y_{bc}A_{cd}Y_{da}+ A_{ab}Y_{bc}Y_{cd}A_{da}
       -A_{ab}X_{bc}^k A_{cd}X_{da}^k \nn
  &&\hspace{17ex}
	+ A_{ab}X_{bc}^k X_{cd}^k A_{da}
	+Y_{ab}X_{bc}^k Y_{cd} X_{da}^k
	-Y_{ab} X_{bc}^k X_{cd}^k Y_{da}\nn
  &&\hspace{17ex}
       +\frac{1}{2}X_{ab}^k X_{bc}^l X_{cd}^k X_{da}^l
       -\frac{1}{2}X_{ab}^k X_{bc}^l X_{cd}^l X_{da}^k\bigg\}\nn
  &&\hspace{1cm}
        +\sum_{a,b,c=1}^N\bigg\{
	i\bar{C}_{ab}A_{ba}C_{ac}A_{ca}
	-i\bar{C}_{ab}A_{ba}A_{ac}C_{ca}
	-iA_{ab}\bar{C}_{ba}C_{ac}A_{ca}\nn
  &&\hspace{15ex}
	+iA_{ab}\bar{C}_{ba}A_{ac}C_{ca}
          -i\bar{C}_{ab}Y_{ba}C_{ac}Y_{ca}
	 +i\bar{C}_{ab}Y_{ba}Y_{ac}C_{ca}\nn
  &&\hspace{15ex}
	+iY_{ab}\bar{C}_{ba}C_{ac}Y_{ca}
	-iY_{ab}\bar{C}_{ba}Y_{ac}C_{ca}
	-i\bar{C}_{ab}X_{ba}^kC_{ac}X_{ca}^k\nn
  &&\hspace{15ex}
	 +i\bar{C}_{ab}X_{ba}^k X_{ac}^kC_{ca}
	 +iX_{ab}^k\bar{C}_{ba}C_{ac}X_{ca}^k
	 -iX_{ab}^k\bar{C}_{ba}X_{ac}^kC_{ca}\bigg\}
	\Bigg],\label{SB2}\\
  \tilde{S}_{1/2}^F&=& L^{1/2}\int d^2 \xi\, \calL_{1/2}^F\nn
  &=& L^{1/2}\int d^2 \xi \Bigg[\sum_{a,b=1}^N\bigg\{
	2\Psi_{ab}^T(\psi_a-\psi_b)A_{ba}\nn
  &&\hspace{18ex}-2\Psi_{ab}^T\gamma^9(\psi_a-\psi_b)Y_{ba}
	-2\Psi_{ab}^T\gamma^k(\psi_a-\psi_b)X_{ba}^k\bigg\}
	\Bigg],\label{SF1/2}\\
  \tilde{S}_{1}^F&=& L\int d^2 \xi\, \calL_{1}^F\nn
  &=& L\int d^2 \xi \Bigg[ \sum_{a,b=1}^N\bigg\{
	i\Psi_{ab}^T\ptau \Psi_{ba}
	-i\Psi_{ab}^T\gamma^9\pth \Psi_{ba}\bigg\}\nonumber
       +\sum_{a,b,c=1}^N\bigg\{
	\Psi_{ab}^T(A_{bc}\Psi_{ca}-\Psi_{bc}A_{ca})\\
  &&\hspace{18ex}
	-\Psi_{ab}^T\gamma^9(Y_{bc}\Psi_{ca}-\Psi_{bc}Y_{ca})
	-\Psi_{ab}^T\gamma^k(X_{bc}^k\Psi_{ca}-\Psi_{bc}X_{ca}^k)
	\bigg\}\Bigg]\label{SF1}.
\end{eqnarray}
The following formulas are useful to obtain the above expressions,
\begin{eqnarray}
  ( [x,X] )_{ab}&=&(x_a-x_b)\,X_{ab}\,,\\
  tr(x[X,Y])&=&tr(X[Y,x])=tr(Y[x,X])=\sum_{a,b=1}^{N}(x_b-x_a)Y_{ab}X_{ba}\,.
\end{eqnarray}



\begin{thebibliography}{99}
\bibitem{HLP}J.~Hughes, J.~Liu and J.~Polchinski,
	``Supermembranes,''
	Phys.\ Lett.\ B {\bf 180}, 370 (1986).

\bibitem{BST}E.~Bergshoeff, E.~Sezgin and P.~K.~Townsend,
	``Supermembranes And Eleven-Dimensional Supergravity,''
	Phys.\ Lett.\ B {\bf 189}, 75 (1987).

\bibitem{DHIS}M.~J.~Duff, P.~S.~Howe, T.~Inami and K.~S.~Stelle,
	``Superstrings In D = 10 From Supermembranes In D = 11,''
	Phys.\ Lett.\ B {\bf 191}, 70 (1987).

\bibitem{Rus}J.~G.~Russo,
	``Supermembrane dynamics from multiple interacting strings,''
	Nucl.\ Phys.\ B {\bf 492}, 205 (1997)
	[arXiv:hep-th/9610018].

\bibitem{SY}Y.~Sekino and T.~Yoneya,
	``From supermembrane to matrix string,''
	Nucl.\ Phys.\ B {\bf 619}, 22 (2001)
	[arXiv:hep-th/0108176].

\bibitem{dHN}B.~de Wit, J.~Hoppe and H.~Nicolai,
	``On The Quantum Mechanics Of Supermembranes,''
	Nucl.\ Phys.\ B {\bf 305}, 545 (1988).

\bibitem{BFSS}T.~Banks, W.~Fischler, S.~H.~Shenker and L.~Susskind,
	``M theory as a matrix model: A conjecture,''
	Phys.\ Rev.\ D {\bf 55}, 5112 (1997)
	[arXiv:hep-th/9610043].

\bibitem{Mot}L.~Motl,
	``Proposals on nonperturbative superstring interactions,''\\
	arXiv:hep-th/9701025.

\bibitem{DVV}R.~Dijkgraaf, E.~Verlinde and H.~Verlinde,
	``Matrix string theory,''
	Nucl.\ Phys.\ B {\bf 500}, 43 (1997)
	[arXiv:hep-th/9703030].

\bibitem{Sus}L.~Susskind,
	``Another conjecture about M(atrix) theory,''
	arXiv:hep-th/9704080.

\bibitem{Sei} N.~Seiberg,
	``Why is the matrix model correct?,''
	Phys.\ Rev.\ Lett.\  {\bf 79}, 3577 (1997)
	[arXiv:hep-th/9710009].

\bibitem{Sen} A.~Sen,
	``D0-branes on $T^n$ and matrix theory,''
	Adv.\ Theor.\ Math.\ Phys.\  {\bf 2}, 51 (1998)
	[arXiv:hep-th/9709220].

\bibitem{Tay} W.~Taylor,
	``D-brane field theory on compact spaces,''
	Phys.\ Lett.\ B {\bf 394}, 283 (1997)
	[arXiv:hep-th/9611042].

\bibitem{HMS} J.~A.~Harvey, G.~W.~Moore and A.~Strominger,
	``Reducing S duality to T duality,''
	Phys.\ Rev.\ D {\bf 52}, 7161 (1995)
	[arXiv:hep-th/9501022].

\bibitem{GHV} S.~B.~Giddings, F.~Hacquebord and H.~Verlinde,
	``High energy scattering and D-pair creation in matrix string
	theory,''
	Nucl.\ Phys.\ B {\bf 537}, 260 (1999)
	[arXiv:hep-th/9804121].

\bibitem{KU} T.~Kugo and S.~Uehara,
	``General Procedure Of Gauge Fixing Based On BRS Invariance
	Principle,''
	Nucl.\ Phys.\ B {\bf 197}, 378 (1982).

\end{thebibliography}
\end{document}